\documentclass[fleqn,usenatbib]{mnras}

\usepackage{newtxtext,newtxmath}

\usepackage[T1]{fontenc}

\DeclareRobustCommand{\VAN}[3]{#2}
\let\VANthebibliography\thebibliography
\def\thebibliography{\DeclareRobustCommand{\VAN}[3]{##3}\VANthebibliography}

\usepackage{graphicx}	
\usepackage{amsmath}	

\title[Making hot Jupiters in stellar clusters]{Making hot Jupiters in stellar clusters II: efficient formation in binary systems}

\author[D. Li et al.]{
Daohai Li,$^{1}$\thanks{E-mail: lidaohai@gmail.com (DL)}
Alexander J. Mustill,$^{2,3}$
Melvyn B. Davies$^{4}$
and  Yan-Xiang Gong$^{5}$\thanks{E-mail: yxgong@tsu.edu.cn (YXG)}
\\
$^{1}$Department of Astronomy, Beijing Normal University, No.19, Xinjiekouwai St, Haidian District, Beijing, 100875, P.R.China\\
$^{2}$Lund Observatory, Department of Astronomy and Theoretical Physics, Lund University, Box 43, 22100 Lund, Sweden\\
$^{3}$Lund Observatory, Division of Astrophysics, Department of Physics, Lund University, Box 43, 22100 Lund, Sweden\\
$^{4}$Centre for Mathematical Sciences, Lund University, Box 118, 221 00 Lund, Sweden\\
$^{5}$College of Physics and Electronic Engineering, Taishan University, Taian 271000, China
}

\date{Accepted XXX. Received YYY; in original form ZZZ}

\pubyear{2015}

\begin{document}
\label{firstpage}
\pagerange{\pageref{firstpage}--\pageref{lastpage}}
\maketitle

\begin{abstract}
Observations suggested that the occurrence rate of hot Jupiters (HJs) in open clusters is largely consistent with the field ($\sim1\%$) but in the binary-rich cluster M67, the rate is $\sim5\%$. How does the cluster environment boost HJ formation via the high-eccentricity tidal migration initiated by the extreme-amplitude von Zeipel-Lidov-Kozai (XZKL) mechanism forced by a companion star? Our analytical treatment shows that the cluster's collective gravitational potential alters the companion's orbit slowly, which may render the star-planet-companion configuration XZKL-favourable, a phenomenon only possible for very wide binaries. We have also performed direct Gyr $N$-body simulations of the star cluster evolution and XZKL of planets' orbit around member stars. We find that an initially-single star may acquire a companion star via stellar scattering and the companion may enable XZKL in the planets' orbit. Planets around an initially-binary star may also be XZKL-activated by the companion. In both scenarios, the companion's orbit has likely been significantly changed by star scattering and the cluster potential before XZKL occurs in the planets' orbits. Across different cluster models, 0.8\%-3\% of the planets orbiting initially-single stars have experienced XZKL while the fraction is 2\%-26\% for initially-binary stars. Notably, the ejection fraction is similar to or appreciably smaller than XZKL. Around a star that is binary at 1 Gyr, 13\%-32\% of its planets have undergone XZKL, and combined with single stars, the overall XZKL fraction is 3\%-21\%, most affected by the cluster binarity. If 10\% of the stars in M67 host a giant planet, our model predicts an HJ occurrence rate of $\sim1\%$. We suggest that HJ surveys target old, high-binarity, not-too-dense open clusters and prioritise wide binaries to maximise HJ yield.
\end{abstract}


\defcitealias{Li2023}{Paper I}
\section{Introduction}\label{sec-int}
Among the over 5000 exoplanets confirmed so far, only a few dozen detections have been made in star clusters \citep[see, e.g.,][for recent compilation and discussion]{Cai2019,Bouma2020}. Small number statistics notwithstanding, it has been suggested that the occurrence rate of exoplanetary systems in open star clusters is not inconsistent with that of the field \citep{Meibom2013}.

Radial velocity surveys for hot Jupiters (HJs) have been made for several open clusters, old and young, small and large \citep{Paulson2004,Quinn2012,Quinn2014,Pasquini2012,Takarada2020}. All clusters seem to have a HJ occurrence rate similar to the field (after metallicity correction) except M67 where 3 HJs were detected and the deduced occurrence rate was 5.6\% around single stars and 4.5\% including binaries also \citep{Brucalassi2016}. This seems to be considerably higher than that of the field \citep[about 1\%, e.g.,][]{Cumming2008,Wright2012}.

Several dedicated works \citep{Shara2016,Hamers2017a,Wang2020a,Rodet2021a,Wang2022,Winter2022a} have looked into the formation of HJs in star clusters as we discussed in \citet[][hereafter \citetalias{Li2023}]{Li2023}. Among these, \citet{Shara2016,Wang2020a,Rodet2021a,Wang2022} examined HJ formation in multi-planet (object) systems whereas \citet{Hamers2017a,Winter2022a} concentrated on single-planet ones. In these latter two publications, the authors showed that consecutive scattering may gradually heat up the planet's orbital eccentricity to very high values and tidal migration may be initiated. However, this mechanism is inefficient and can reach an efficiency of a few per cent at most. We note that their most favourable stellar density for HJ formation is typical of globular clusters, but not of open clusters.

For an open cluster with non-zero binarity, \citet{Li2020c} showed that a member planetary system may while remaining intact, obtain a companion via scattering with a binary. \citetalias{Li2023} examined the effect of the companion and found it may be able to excite large amplitude oscillations in the planet's orbit, giving rise to tidal circularisation and shrinkage of the orbit and HJ formation. The efficiency of this process as derived in \citetalias{Li2023} is $\lesssim1\%$ for M67.

However, the setup of \citetalias{Li2023} was idealised in that their cluster parameters were stationary (for instance, they used a fixed stellar density). Moreover, only singles have planets in \citetalias{Li2023} and therefore planets formed in binaries have been completely overlooked. Furthermore, \citetalias{Li2023} only considered the effect of scattering with stars very close to the planetary system but not the other cluster member stars that are distant but whose combined gravitational pull still exerts a tidal effect. In this work, we extend \citetalias{Li2023} to investigate how these factors influence HJ formation in star clusters.

Before presenting the work, we would like to clarify the acronyms/conventions adopted in this paper. Firstly, ``ZKL'' is short for the von Zeipel-Lidov-Kozai mechanism \citep{Zeipel1909,Kozai1962,Lidov1962}. Because the tidal interaction is not explicitly included here, we are not able to study directly the formation of an HJ so we simply count planets that ever have their pericentre distances smaller than 0.022 au \citep{Beauge2012}. Those planets may become HJs if the tides are strong or otherwise plunge into the host star. We refer to those planets as ``XZKL''(extreme-amplitude ZKL)-activated. In this work, we are only interested in HJ formation around solar mass stars with a mass in the range $(0.8,1.2)$ solar mass ($\mathrm{M}_\odot$). In the numerical simulations, these objects may be single stars or binary stars; and the two statuses may be referred to at the beginning or the end of the simulations. We call these initially/finally-single and initially/finally-binary, respectively.

The paper is organised as follows. In Section \ref{sec-pot}, we study the formation of XZKL due to the perturbation of a companion star that is itself perturbed by the cluster potential approximated by simple analytical terms. In Section \ref{sec-pop}, we track the evolution of planet-hosting stars in star clusters using $N$-body simulations (so both star scattering and cluster potential have been accounted for) and later use the information to model XZKL formation around those stars. The implications are discussed in Section \ref{sec-dis} and the results are presented in Section \ref{sec-con}, respectively.

\section{Effect of cluster potential on planet's XZKL in binaries}\label{sec-pot}
When residing in a star cluster, a stellar binary (a member hosting a planet) feels not only the gravitational forces of stars in its immediate vicinity but also those farther away, both effects able to alter the binary relative motion. The former may modify the companion's orbit abruptly and has been shown to substantially boost the fraction of planets experiencing XZKL around the component stars (\citetalias{Li2023}). The vast majority of the member stars of the cluster, while relatively weak individually, also contribute to the binary relative orbital evolution and their cumulative effect can be approximated by the cluster potential.

\begin{figure}
\includegraphics[width=\columnwidth]{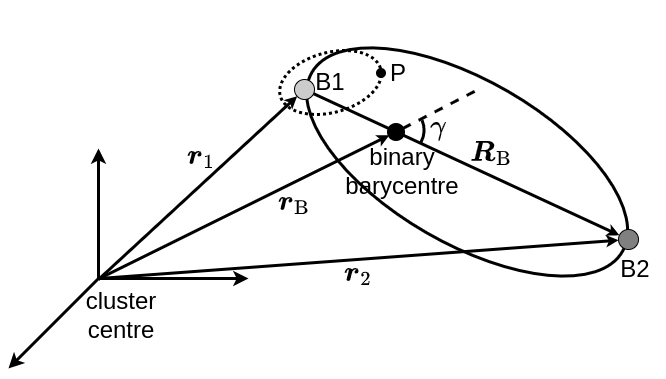}
\caption{Illustration of the setup of the cluster-binary-planet system model. A planet (labelled ``P'', small black point) is orbiting a star B1 (light grey point) which is accompanied by star B2 (dark grey point). The two stars B1 and B2 revolve around their barycentre (black point) and are a member of a star cluster of which the centre is at the origin. The vectors $\boldsymbol r_\mathrm{B}$, $\boldsymbol r_1$, and $\boldsymbol r_2$ measure the locations of the barycentre, B1, and B2. The vector $\boldsymbol R_\mathrm{B}$ points towards B2 from B1 and the angle between $\boldsymbol R_\mathrm{B}$ and $\boldsymbol r_\mathrm{B}$ is $\gamma$. The figure is not to scale in that in actuality $r_\mathrm{B}\gg R_\mathrm{B}$.}
\label{fig-pot-illu}
\end{figure}

In this section, we analytically study how the cluster potential alters the binary relative orbit, and whether and how this affects XZKL in the planet's orbit. As shown in Figure \ref{fig-pot-illu}, we assume a stellar binary comprising B1 and B2, the former orbited by a planet. The planet is treated as a massless test particle throughout the paper since its mass is much smaller. The binary barycentre is moving in a star cluster which is modelled using potential theories.

\subsection{Secular theories}\label{sec-pot-ana}
A number of analytical models have been proposed for the cluster potential \citep{Binney2008}. Here we follow the widely-used Plummer model \citep{Plummer1911} for its simplicity. The potential of the Plummer model at a point takes the form \citep[e.g.,][]{Binney2008}
\begin{equation}
\label{eq-cluster-pot}
\phi=-{GM \over \sqrt{r^2+b^2}},
\end{equation}
where $M$ is the cluster's total mass, $b$ the Plummer scale length, and $r$ the distance of the point from the cluster centre. Both binary components feel the cluster potential. The total potential energy of the cluster-binary-planet system is thus (the planet is massless)
\begin{equation}
\label{eq-tot-eng}
\Phi=-{Gm_1m_2\over R_\mathrm{B}}-{GMm_1 \over \sqrt{r_1^2+b^2}}-{GMm_2 \over \sqrt{r_2^2+b^2}},
\end{equation}
where $r_1$ and $r_2$ are the cluster-centric distances of the two stars and $R_\mathrm{B}$ their mutual distance. The first term on the right-hand side is the two components' mutual gravitational potential energy. The last two terms are the potential energy of the two stars related to the cluster. Letting $r_\mathrm{B}$ be the distance from the binary centre of mass to the cluster centre, we have $|{\boldsymbol r}_1-{\boldsymbol r}_\mathrm{B}|\sim|{\boldsymbol r}_2-{\boldsymbol r}_\mathrm{B}|\sim R_\mathrm{B}$. We also note similar to $b$, $r_\mathrm{B}$ is of order pc, while $R_\mathrm{B}$ (also the semimajor axis of the binary relative orbit $a_\mathrm{B}$) is under a few thousand of au. Thus the ratio $a_\mathrm{B}/r_\mathrm{B}$ is a small quantity and Equation \eqref{eq-tot-eng} can be expanded in it. Moreover, for $a_\mathrm{B}$ as wide as a few thousand of au, the corresponding orbital period is no longer than a few tens of thousands of years, much shorter than the timescale of the evolution of the orbit itself (the precession of the ascending node for instance). So the binary's relative orbital motion can be averaged out from the above potential energy.

The resultant (singly-)averaged potential energy between the binary orbit and the cluster in the vector form, at the quadrupole level in $a_\mathrm{B}/r_\mathrm{B}$, is
\begin{equation}
\label{eq-bin-pot}
\phi_\mathrm{B}=-C_\mathrm{B}[1-6e_\mathrm{B}^2-(2+3e_\mathrm{B}^2){b^2\over r_\mathrm{B}^2}+15({\boldsymbol e}_\mathrm{B} \cdot \hat r_\mathrm{B})^2-3({\boldsymbol j}_\mathrm{B} \cdot \hat r_\mathrm{B})^2],
\end{equation}
where the coefficient
\begin{equation}
C_\mathrm{B}={GMm_1m_2 r_\mathrm{B}^2a_\mathrm{B}^2\over 4(m_1+m_2)(r_\mathrm{B}^2+b^2)^{5/2}}.
\end{equation}
Here $a_\mathrm{B}$ and $e_\mathrm{B}$ are the semimajor axis and eccentricity of the binary relative orbit; ${\boldsymbol e}_\mathrm{B}$ is the vector pointing towards the pericentre of the binary relative orbit with a length of $e_\mathrm{B}$; the vector ${\boldsymbol j}_\mathrm{B}$ is parallel to the orbital normal of the binary relative motion and has a length of $\sqrt{1-e_\mathrm{B}^2}$; $\hat r_\mathrm{B}$ is the unit vector pointing towards the barycentre of the binary. The details of the expansion and averaging are given in the appendix. We note that when the $b/r$ is small, the above equation reduces to the point mass case \citep{Tremaine2014}.

Then the system total potential energy \eqref{eq-tot-eng} becomes
\begin{equation}
\label{eq-tot-eng-2}
\Phi=-{Gm_1m_2\over R_\mathrm{B}}-{GM \over \sqrt{r_\mathrm{B}^2+b^2}}+\phi_\mathrm{B}.
\end{equation}
The first term on the right-hand side dictates the two binary components' relative motion, the second the motion of the binary barycentre in the cluster, and the third the evolution of the binary relative orbit. The equations of motion for ${\boldsymbol e}_\mathrm{B}$ and ${\boldsymbol j}_\mathrm{B}$ under the potential \eqref{eq-bin-pot} are given in the appendix (Equations \eqref{eq-eq-eb} and  \eqref{eq-eq-jb}).

In the above, only the inner orbital motion (the binary relative motion) has been averaged out. One might ask if the outer orbital motion (that of the binary barycentre) can be also eliminated. One problem is that the trajectory of the barycentre is often not closed and there is no simple analytical expression to describe it \citep{Heggie2003,Binney2008}. A trivial case is where the binary centre of mass is moving on a circular orbit around the cluster centre. Then, $r_\mathrm{B}$ is a constant and $\hat r_\mathrm{B}$ rotates in a plane evenly with time (because the cluster potential is symmetric) and a further averaging can be done for $\hat r_\mathrm{B}$ (on condition that the period of the motion of the binary barycentre is much shorter than the evolution of the binary relative orbit, which we discuss in Section \ref{sec-pot-time}). The resulting potential expressed in the vector form is Equation \eqref{eq-bin-pot-ave} in the appendix. Here we present the same potential written in the usual orbital elements
\begin{equation}
\label{eq-bin-pot-ave-ele}
\begin{aligned}
\phi_\mathrm{B}={C_\mathrm{B}\over 2}[1&-6e_\mathrm{B}^2+{b^2\over r_\mathrm{B}^2}(4+6e_\mathrm{B}^2)\\
+&15e_\mathrm{B}^2\sin^2\omega_\mathrm{B}\sin^2i_\mathrm{B}-3(1-e^2_\mathrm{B}) \cos^2 i_\mathrm{B}],
\end{aligned}
\end{equation}
where $i_\mathrm{B}$, $\omega_\mathrm{B}$ are the inclination and argument of pericentre of the binary's relative orbit, measured against the plane of motion of their barycentre in the cluster (which is stationary because of the symmetry of the cluster potential). This expression is again very much like that of the standard quadrupole ZKL formalism (cf. Equation \eqref{eq-pla-quad}) except the addition of the term ${b^2\over r_\mathrm{B}^2}(4+6e_\mathrm{B}^2)$.

\begin{figure}
\includegraphics[width=\columnwidth]{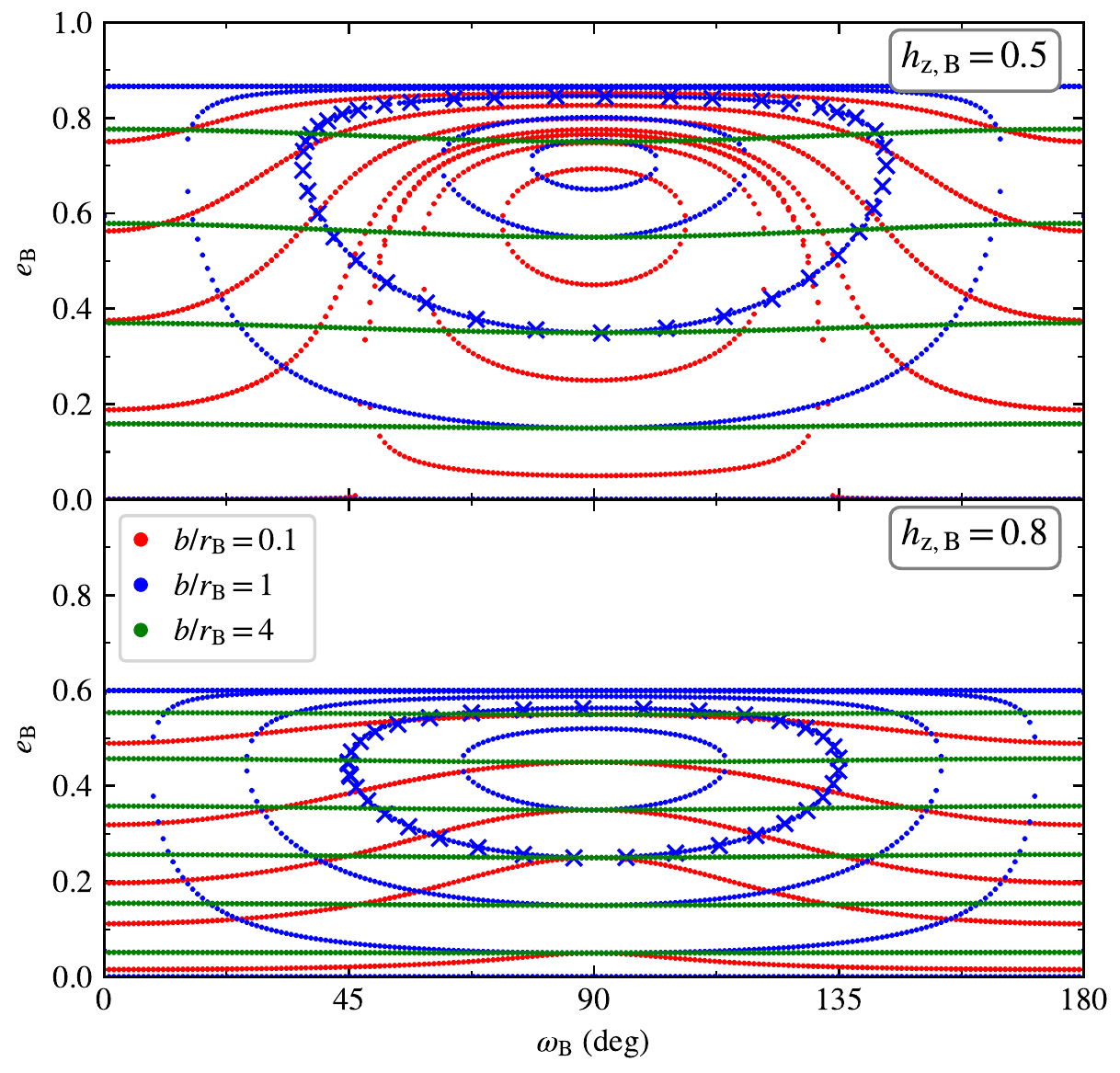}
\caption{Phase diagram of the system described by Equation \eqref{eq-bin-pot-ave-ele}. In the top panel, $h_\mathrm{z,B}=0.5$ and in the bottom $h_\mathrm{z,B}=0.8$. Red, blue and green points represent different constant $b/r_\mathrm{B}=0.1$, 1, and 4, respectively; this is also the reason why there is overlap between the level curves of different colours. The crosses are from $N$-body simulations with the same energy.}
\label{fig-phase}
\end{figure}

Because of the absence of the longitude of ascending node $\Omega_\mathrm{B}$ from Equation \eqref{eq-bin-pot-ave-ele}, its conjugate angular momentum $h_\mathrm{z,B}=\sqrt{1-e_\mathrm{B}^2}\cos i_\mathrm{B}$ is conserved. Therefore, the system \eqref{eq-bin-pot-ave-ele} is integrable. In Figure \ref{fig-phase} we plot the phase diagram in the $\omega_\mathrm{B}-e_\mathrm{B}$ space for two values of $h_\mathrm{z,B}$.

In the top panel, $h_\mathrm{z,B}=0.5$ and level curves for three values of $b/r_\mathrm{B}$ have been plotted. When $b/r_\mathrm{B}$ is small (0.1, red points), the curves are much like those in the quadrupole ZKL \citep{Kozai1962,Ford2000} and there is a libration centre at $\omega_\mathrm{B}=90^\circ$. For an intermediate value of $b/r_\mathrm{B}$ (1, blue points), the libration region seems to be larger so the variation in $e_\mathrm{B}$ may be larger. When $b/r_\mathrm{B}=4$, the libration region disappears.

In the bottom panel ($h_\mathrm{z,B}=0.8$), it seems that only for $b/r_\mathrm{B}=1$, there appears a libration region; for smaller and larger values of $b/r_\mathrm{B}=4$, only circulation is possible and the change is $e_\mathrm{B}$ is then relatively small.

In order to validate the secular theory, we have also explored the evolution of the stellar binary using our modified $N$-body package {\small MERCURY} \citep{Chambers1999}. We have added to the code the capability of tracking both the motion of the binary barycentre in the cluster and that of one binary component relative to the other under the Plummer potential. In the two panels of Figure \ref{fig-phase}, we have for each, selected a random energy level and showed the evolution of the binary relative orbit using the $N$-body method as outlined above using the crosses. The agreement between the $N$-body simulation and the secular theory is excellent.

The planet's orbital evolution around the star B1, on the other hand, is governed by the companion star B2. The potential in the vector form, averaged over both the planet's and companion's orbital periods and truncated to the quadrupole level in the ratio $a/a_\mathrm{B}$, is \citep{Liu2015,Petrovich2015}
\begin{equation}
\label{eq-pla-quad}
\phi_\mathrm{P}={Gm_2a^2\over 8a_\mathrm{B}^3(1-e_\mathrm{B}^2)^{5/2}}[(1-6e^2)(1-e_\mathrm{B}^2)-3({\boldsymbol j}\cdot {\boldsymbol j}_\mathrm{B})^2+15({\boldsymbol e}\cdot {\boldsymbol j}_\mathrm{B})^2]
\end{equation}
where $a$, $e$, ${\boldsymbol e}$, and ${\boldsymbol j}$ are the respective quantities for the planet's orbit and we have also let the planet's mass be zero. Higher order expansions \citep{Ford2000,Naoz2013} are not included as the purpose is to qualitatively showcase how the cluster potential can modulate the ZKL cycles in the planet by altering the companion's orbit and the leading order theory suffices. The equations of motion of the planet's orbital evolution are given in the Appendix and can also be found in \citet{Liu2015,Petrovich2015}. We call the dynamics encompassed by this potential the standard ZKL mechanism at quadrupole level.

Now it can be readily seen why adopting the vector-form expressions for the potentials is sensible. Otherwise, when expressed in orbital elements, the system's total angular momentum sets the reference plane \citep[e.g.,][]{Valtonen2006,Naoz2013}, i.e., the orbital plane of the binary relative motion. But this plane is itself evolving over time because of the cluster potential, in which case a reference frame conversion is needed.
\subsection{Possible effect on HJ formation}\label{sec-pot-num}
Above we have detailed our secular description of the cluster-binary-planet system as illustrated in Figure \ref{fig-pot-illu} -- the binary barycentre moves according to the middle term of the right-hand side of Equation \eqref{eq-tot-eng-2} and the binary relative orbit is dictated by Equation \eqref{eq-bin-pot}; these two types of evolution are controlled by the cluster potential. The planet's orbit, on the other hand, is governed by the companion's perturbation \eqref{eq-pla-quad}. This secular system can be numerically solved.

\begin{figure}
\includegraphics[width=\columnwidth]{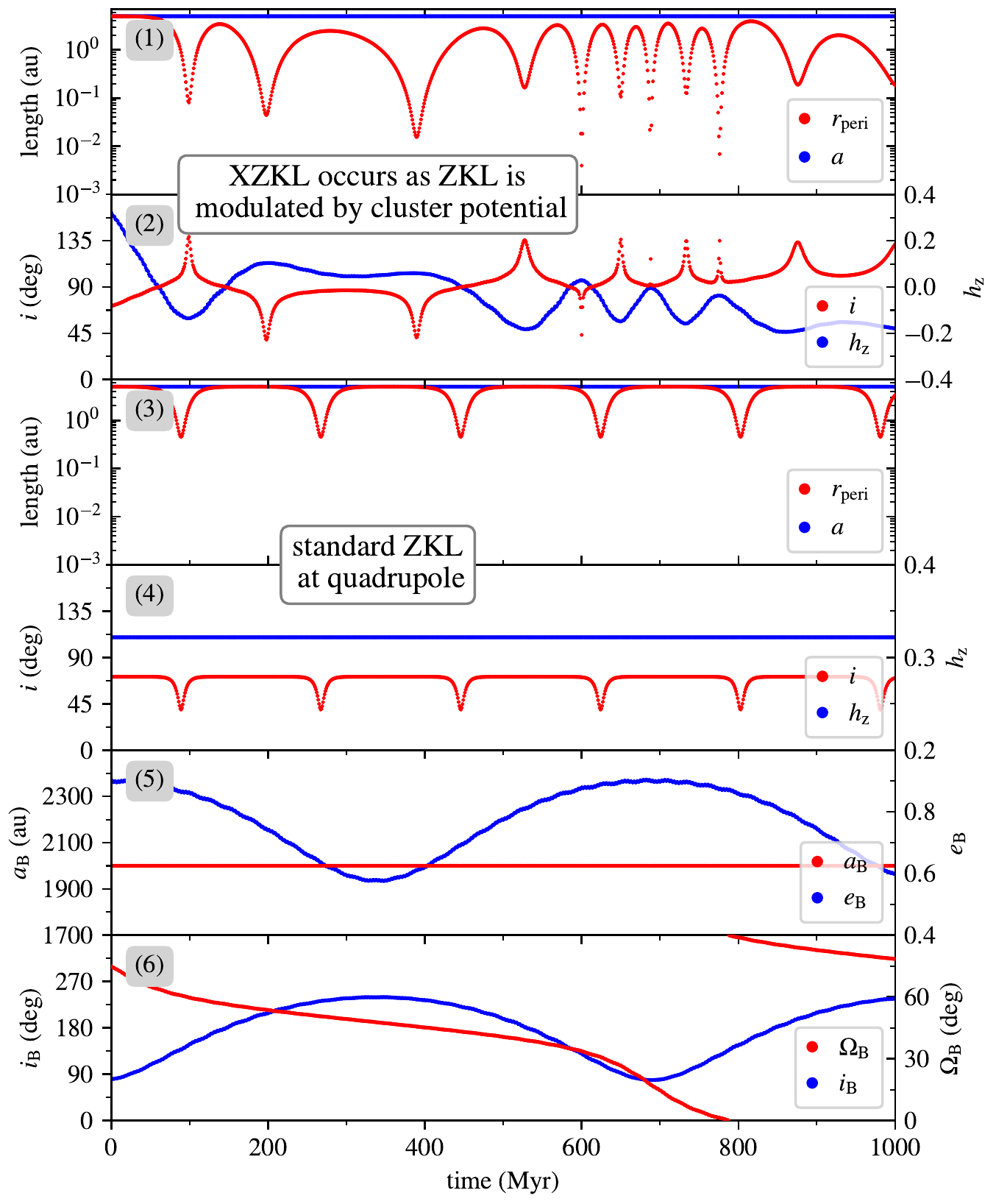}
\caption{Time evolution of the XZKL process in the secular model. Panel (1) shows the pericentre distance of the planet's orbit $r_\mathrm{peri}$ (red) and the semimajor axis $a$ (blue); (2) shows the planet's inclination $i$ (red, left $y$-axis) and $h_\mathrm{z}=\sqrt{1-e^2} \cos i$ (blue, right $y$-axis); the reference plane is that of the binary relative motion; the ZKL cycles have been modulated by cluster's potential by altering the companion orbit (Panels (5) and (6)) and XZKL results. Panels (3) and (4) show the same planet, the plot arrangement the same as above and the difference being that the cluster potential has been turned off; XZKL never occurs. Panel (5) shows the companion star's relative semimajor axis $a_\mathrm{B}$ (red, left $y$-axis) and eccentricity $e_\mathrm{B}$ (blue, right $y$-axis) and (6) shows the longitude of ascending node $\Omega_\mathrm{B}$ (red, left $y$-axis) and inclination $i_\mathrm{B}$ (blue, right $y$-axis); the reference plane is that of the motion of the binary centre of mass.}
\label{fig-analytical}
\end{figure}

In Figure \ref{fig-analytical} we show an example where XZKL has been activated in the planet orbit, which is not possible without the cluster potential. We emphasise that as discussed later in Section \ref{sec-pot-time} enabling XZKL by the cluster potential may not be the norm and this example only serves to highlight the possibility.

The initial setup of the cluster of the underlying simulation of Figure \ref{fig-analytical} is $M=1000\mathrm{M}_\odot$ and $r_\mathrm{h}=2$ pc. The binary properties are $m_1=1\mathrm{M}_\odot$, $m_2=0.5\mathrm{M}_\odot$; the relative orbit is $a_\mathrm{B}=2000$ au, $e_\mathrm{B}=0.9$, $i_\mathrm{B}=20^\circ$, $\omega_\mathrm{B}=100^\circ$, $\Omega_\mathrm{B}=300^\circ$; and the initial motion of barycentre is $r_\mathrm{B}=1$ pc, the velocity vector ${\boldsymbol v}_\mathrm{B}\perp \hat r_\mathrm{B}$ and the magnitude is 1.2 times $v_\mathrm{B,cir}$ (the circular velocity at $r_\mathrm{B}$). The planet parameters are $a=5$ au, $e=0.01$, $i=60^\circ$, $\omega=200^\circ$, $\Omega=60^\circ$. We note when describing the initial condition, the planet's orbit is measured against the plane of motion of the binary barycentre; wherever else, the reference plane has been that of the binary relative motion when discussing the planet's orbit throughout the paper (including Figure \ref{fig-analytical}). The resulting planet's initial inclination with respect to the binary relation orbital plane is $71^\circ$. Our choice of the initial condition is arbitrary but shows the effect of the cluster potential.

Panels (5) and (6) of Figure \ref{fig-analytical} show the evolution of the binary. We observe that $a_\mathrm{B}$ is constant, a result of orbital averaging, and $e_\mathrm{B}$ oscillates between 0.6 and 0.9 with a period of $\sim 700$ Myr. Then the phase angle $\Omega_\mathrm{B}$ circulates, barely finishing a cycle in 1 Gyr; the inclination $i_\mathrm{B}$ varies between $20^\circ$ and $60^\circ$, in an anti-correlated phase with $e_\mathrm{B}$. The cluster-centric distance of the binary barycentre $r_\mathrm{B}$ roughly oscillates between 1 and 1.3 pc on a much shorter timescale and is not shown.

Panels (1) and (2) show the evolution of the planet perturbed by the companion. In Panel (1), the planet's $a$ is constant while the pericentre $r_\mathrm{peri}$ shows two types of variation: one on a timescale $\lesssim 100$ Myr and the other (the extrema of $r_\mathrm{peri}$) on a longer timescale. We note the minimum reaches 0.015 au at 400 Myr and 0.004 au at 600 Myr, where tides (if modelled) may give rise to the formation of an HJ or collision between the planet and the central host (or  tidal disruption of the planet) may result. In either case, XZKL occurs. The inclination $i$ evolves largely between $40^\circ$ and $140^\circ$, in correlation with $r_\mathrm{peri}$, and thus the planet's trajectory alternates between prograde and retrograde statuses. In the standard ZKL, the planet's vertical orbital angular momentum $h_\mathrm{z}=\sqrt{1-e^2} \cos i$ is conserved. Herein, because of the evolution of the companion's orbit seen in the bottom two panels, $h_\mathrm{z}$, from an initial value of 0.32, fluctuates; the value of $h_\mathrm{z}$ approaches/crosses zero a few times, where orbital flipping happens and XZKL is possible.

Then Panels (3) and (4) show the evolution of the same planet, now the cluster potential switched off (so the binary relative orbit is not changing). The planet's orbital evolution can be described by the standard ZKL and now $r_\mathrm{peri}$ and $i$ oscillate but never achieve extreme values and XZKL never occurs.

Though the example in Figure \ref{fig-analytical} clearly shows that modulation in the binary relative orbit by the cluster potential may help with the planet's XZKL, we remind again that this scenario may not be typical and is only to showcase the mechanism. As we discuss below, caution needs to be taken.
\subsection{Timescale comparison}\label{sec-pot-time}
Several types of motion and timescales are involved in our system of cluster-binary-planet (Figure \ref{fig-pot-illu}). In this section we briefly discuss them.

The centre of mass of the binary moves in the cluster and its timescale is usually referred to as the crossing time $\tau_\mathrm{cross}$. Here we approximate it as $\tau_\mathrm{cross}=r_\mathrm{B}/v_\mathrm{B,cir}$ and $v_\mathrm{B,cir}$ can be solved for by equating the centripetal acceleration with the gravitational acceleration of the cluster which is derived from the Plummer potential.

As established in Section \ref{sec-pot-ana}, the cluster potential also causes the binary's relative orbit to evolve. Here we estimate its timescale using $\tau_\mathrm{\Omega,B}$, the precession timescale of $\Omega_\mathrm{B}$. And $\tau_\mathrm{\Omega,B}$ can be readily derived using the doubly-average potential energy \eqref{eq-bin-pot-ave-ele} and Lagrange's planetary equations \citep[e.g.,][]{Valtonen2006}. The resultant expression is
\begin{equation}
\label{eq-tau-B}
\begin{aligned}
\tau_\mathrm{\Omega,B}&={1\over\dot \Omega_\mathrm{B}}={L_\mathrm{B} \sqrt{1-e^2_\mathrm{B}}\sin i_\mathrm{B}\over \partial \phi_\mathrm{B}/ \partial i_\mathrm{B}}\\
&={L_\mathrm{B} \sqrt{1-e^2_\mathrm{B}} \over C_\mathrm{B}[15e_\mathrm{B}^2\sin^2\omega_\mathrm{B}+3(1-e^2_\mathrm{B})]\cos i_\mathrm{B}}\\
&\approx {4 L_\mathrm{B} \sqrt{1-e^2_\mathrm{B}}\over3 C_\mathrm{B}(1+3e^2_\mathrm{B})},
\end{aligned}
\end{equation}
where $L_\mathrm{B}$ is one of the Delaunay elements \citep[see also][]{Murray1999} as defined in Equation \eqref{eq-lb} in the appendix. The term $\sin^2\omega_\mathrm{B}$ in the middle line has been ``averaged'' in the third line assuming that $\omega_\mathrm{B}$ distributes evenly in the range $(0,360)$ deg; and the term $\cos i_\mathrm{B}$ is also ``averaged'' assuming that the orbital pole is isotropically distributed in space (so the probability density function of $\cos i_\mathrm{B}$ is flat in the range $(0,1)$). We also note that compared to the original Lagrange's formulae \citep{Murray1999,Valtonen2006}, ours is slightly non-standard as a binary of two massive bodies is being examined as opposed to a massless object around a star.

Then, the numerous weak encounters with distant stars may deflect the motion of the binary centre of mass slowly. Over time, this may change the trajectory of the binary barycentre significantly, a process called relaxation. Here we use the cluster's half-mass relaxation time $\tau_\mathrm{r,h}$ \citep{Spitzer1987}, and take the expression from \citet{Heggie2003}
\begin{equation}
\tau_\mathrm{r,h}={0.206Nb^{3/2}\over \sqrt{GM}\ln \Lambda},
\end{equation}
where $N$ in the number of stellar systems and we assume an average star is $0.5 \mathrm{M}_\odot$, so $N$ can be obtained knowing the cluster total mass; $\Lambda$ is the Coulomb logarithm and takes the value $0.4N$ in our evaluation \citep{Spitzer1987}.

Furthermore, neighbouring stars may scatter strongly with the binary, changing the relative orbit in an instant. Here we consider the binary's dissociation, of which the timescale can be estimated using 
\begin{equation}
\tau_\mathrm{com}={1\over n\sigma v},
\end{equation}
where $n$ is the stellar density, $\sigma$ the corresponding cross-sectional areas \citep{Hut1983,Bacon1996}, and $v$ (assumed to be 0.5 km s$^{-1}$ here) is the encountering velocity at infinity. We assume that any encountering star is 0.5 $\mathrm{M}_\odot$ in our calculation.

Finally, the planet's orbital evolution is driven by the companion as encompassed by Equation \eqref{eq-pla-quad}. Its timescale $\tau_\mathrm{ZKL}$ can be estimated following \citet{Antognini2015}
\begin{equation}
\tau_\mathrm{ZKL}={8\over 15\pi}{m_1+m_2\over m_2}{P^2_\mathrm{B} \over P}(1-e^2_\mathrm{B})^{3/2},
\end{equation}
where $P_\mathrm{B}$ is the orbital period of the binary relative motion and $P$ that of the planet.

\begin{figure*}
\includegraphics[width=1.5\columnwidth]{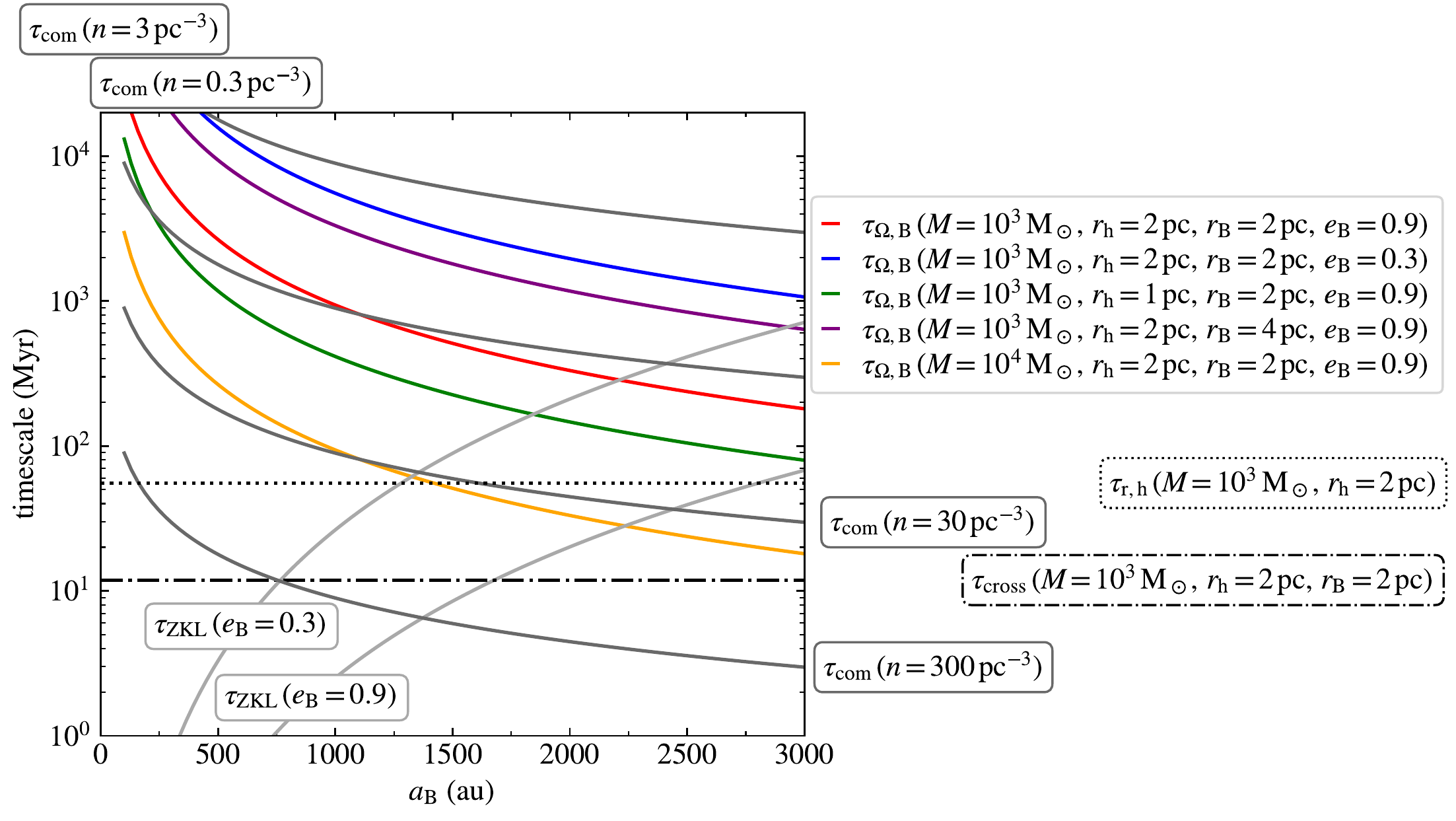}
\caption{Relevant timescales as a function of the binary relative semimajor axis $a_\mathrm{B}$. The coloured lines show $\tau_\mathrm{\Omega,B}$ that of the precession of the relative orbit caused by the cluster potential for different cluster and binary parameters as indicated by the legends. The dark grey lines show $\tau_\mathrm{com}$ the dissociation timescale of the binary due to stellar scattering for different stellar densities. The light grey lines show $\tau_\mathrm{ZKL}$ the ZKL timescale of the planet's orbit for different $e_\mathrm{B}$. The black dash-dotted horizontal and the black dotted horizontal lines show $\tau_\mathrm{cross}$ the crossing time and $\tau_\mathrm{r,h}$ the half-mass relaxation time of the cluster, respectively.}
\label{fig-timescale}
\end{figure*}

In Figure \ref{fig-timescale}, we compare these timescales as a function of $a_\mathrm{B}$. The binary masses are $m_1=1\mathrm{M}_\odot$, $m_2=0.5\mathrm{M}_\odot$ and their barycentre moves on a circular path of radius $r_\mathrm{B}$ in the cluster. The planet is around its host star in a circular orbit at $a=5$ au. The other parameters are varied, as labelled in the plot. 

In general, the value of $\tau_\mathrm{\Omega,B}$ (coloured lines) spans more than two orders of magnitude from a few tens of Myr for a very wide and eccentric binary in a massive cluster to $\gtrsim 10^4$ Myr for a tight, less eccentric binary in a relatively small cluster. The negative dependence on $a_\mathrm{B}$ is obvious: for a wider binary, the components feel more distinct cluster potentials, causing faster precession in the relative orbit. The companion lifetime $\tau_\mathrm{com}$, shown by the dark grey lines, is anti-correlated with $a_\mathrm{B}$. Wide binaries $\gtrsim1000$ au cannot survive for more than 10 Myr in dense 300 pc$^{-1}$ clusters but those tight $\lesssim$ 400 au can remain stable over the age of the universe in sparsely-populated 0.3 pc$^{-1}$ clusters. The ZKL timescale $\tau_\mathrm{ZKL}$ depends on $a_\mathrm{B}$ positively and $e_\mathrm{B}$ also plays an important role. Finally, the quantities $\tau_\mathrm{cross}$ and $\tau_\mathrm{r,h}$ do not depend on $a_\mathrm{B}$ and are shown by the horizontal dash-dotted and dotted black lines. The two are also not very sensitive to the cluster parameters, varying by a factor of only several across the parameters used for $\tau_\mathrm{\Omega,B}$, so we only plot one for each.

What is apparent in Figure \ref{fig-timescale} is that the timescales may not be well separated and therefore the dynamics may be complicated. This casts doubt on the validity of the secular theory presented above. For instance, when deriving $\tau_\mathrm{\Omega,B}$, the simplification that the barycentre path is a circle of radius $r_\mathrm{B}$ has been made. However, if $\tau_\mathrm{cross}>\tau_\mathrm{\Omega,B}$ (possible for very wide binaries in massive clusters), the orbital averaging over the motion of the binary barycentre simply cannot be made. In addition, when the binary is wide and the cluster is dense, it is possible that $\tau_\mathrm{r,h}<\tau_\mathrm{\Omega,B}$ or $\tau_\mathrm{com}<\tau_\mathrm{\Omega,B}$; but the change in the trajectory of the motion the binary centre of mass and the dissociation of the binary have not been taken into consideration in the secular treatment. Considering these, we are not delving into the details of the secular description of the system.

Yet, even if in the parameter space where our secular model fails, the binary relative motion is still subject to orbital alternation by the cluster potential so the planet's ZKL cycles might be affected. Whether and how this mechanism works in a cluster environment can only be accessed in full-fledged numerical simulations.

\section{Population synthesis}\label{sec-pop}
In Sections \ref{sec-pot-ana} and \ref{sec-pot-num}, we have used simple analytical potential theory in the cluster-binary-planet setup and showed that the orientation of the companion orbit can be changed by the cluster potential and this may cause XZKL, otherwise impossible, in the planet's orbit. But the timescale analysis in Section \ref{sec-pot-time} suggests that the applicability of this theory is possibly limited to wide binaries in not densely populated clusters. And the mechanism, even if functioning, is likely affected by, for example, scattering, relaxation, and the cluster's overall evolution. These need to be tackled using $N$-body simulations of the star cluster.

In this section, we study the XZKL in a more realistic setup. First in Section \ref{sec-clu}, we numerically track the evolution of planet-hosting stars in the parent clusters during which information of neighbour/companion stars that may affect the planet's ZKL is recorded. This is then passed on to Section \ref{sec-pla} where the ZKL phenomenon of the planets' orbits is probed.
\subsection{Cluster simulation}\label{sec-clu}

\subsubsection{Simulation methods}
Our star cluster simulations are carried out using {\small NBODY6++} \citep{Spurzem1999,Aarseth2003,Wang2015}. The package {\small McLuster} \citep{Kupper2011} is used to create the initial conditions. For each star, its mass is drawn from the Kroupa initial mass function \citep[IMF,][]{Kroupa2001}. The semimajor axis of the binary relative orbit takes a lognormal distribution following that of the field \citep{Duquennoy1991,Raghavan2010} and for the eccentricity, a flat distribution has been assumed; the masses of the two components are drawn independently from the IMF. The initial distribution of the stars follows the Plummer model \citep{Plummer1911,Binney2008}. The single and binary stars are evolved using the prescription of \citet{Hurley2000,Hurley2002}, where a metallicity of 0.02 is used. The cluster feels the solar neighbour galactic tidal field.

The other cluster parameters are varied in the different simulations as listed in Table \ref{tab-pop}. In short, we have seven parameter sets. For the NomiClu set, a cluster contains 2000 stellar systems and the half mass radius $r_\mathrm{h}$ is 2 pc. The binarity $f_\mathrm{b}$ is 50\% meaning that 1000 stellar systems are binary. A total of 20 runs are performed for this parameter choice with different random number seeds. In the LowDen and HighDen clusters, $r_\mathrm{h}$ is set to 4 and 1 pc, respectively; 5 runs are done for each. Then for the LowBin and HighBin sets, $f_\mathrm{b}$ is 10\% and 90\%, respectively. Next, the LargClu set represents a larger cluster with 20000 stellar systems. All these systems assume initial virial equilibrium and are tracked for 1 Gyr. We note the evolution timescales of these cluster models may be vastly different. For instance, the initial half-mass relaxation timescales vary by more than an order of magnitude in those different clusters so they may not be evolved to the same degree when the simulation is terminated. Finally, we also set up a Field run, where the single/binary stars are created by {\small McLuster} but each binary is thought to be in isolation so no cluster simulation is needed and the initial companion information is directly sent to the planets' XZKL simulations. This is effectively simulating XZKL in the field \citep[through the galactic potential and field star scatterings are omitted, cf.][]{Kaib2013}
\begin{table}
\centering
\caption{Initial setup of the cluster simulations. The first column is the identity of the parameter set, the second the number of star systems $N$, the third the half-mass radius $r_\mathrm{h}$, the fourth the binarity $f_\mathrm{bin}$, and the fifth the number of runs with different random number seeds run\_cnt.  The last row represents the case of field stars, so this set is not actually integrated in the cluster simulation.}
\label{tab-pop}
\begin{tabular}{cccccc} 
\hline
sim ID &$N$& $r_{h}$ (pc) &$f_\mathrm{bin}$ &run\_cnt\\
\hline
NomiClu & 2000 & 2 & 50\%&  20\\
LowDen & 2000 & 4 & 50\%&  5\\
HighDen & 2000 & 1 & 50\%&  5\\
LowBin & 2000 & 2 & 10\%&  5\\
HighBin & 2000 & 2 & 90\%&  5\\
LargClu & 20000 & 2 & 50\%& 1\\
Field & 10000 & $\infty$ & 50\%& 1\\
\hline
\end{tabular}
\end{table}

The planets are not included in the cluster simulations but are studied in another set of integrations as detailed in Section \ref{sec-pla}. A number of works have explored the evolution of planetary systems in star clusters adopting this cluster-then-planet approach \citep[e.g.,][]{Cai2017,Vincke2018,Fujii2019}. Usually in the first step, the encounter history of planet-hosting stars are stored and later used in the second. Here in our case, because the binaries are of special interest, we record not only the neighbouring scattering stars but also any companion star on bound orbit around planet-hosting stars in the cluster simulation. To be exact, we look for the closest neighbour of every solar mass star every $\sim 5\times10^2$ yr: if the neighbour is a passerby and its distance is smaller than 2000 au or if the neighbour is bound and any of the four orbital elements $a_\mathrm{B}$, $e_\mathrm{B}$, $i_\mathrm{B}$, $\Omega_\mathrm{B}$ has changed by more than 2\% since the last record, the mass and the 3D relative position and velocity vectors of that star are stored. And a forced output will be implemented if no output has been done for ten consecutive checks so the maximum output interval is $\sim 5\times10^3$ yr. Following \citet{Curtis2014}, we estimate that it takes an encountering star thousands of years to enter and leave a sphere of a radius of 1000 au around the planet-hosting star should that star's pericentre distance $<$ 900 au; so it is unlikely that any potentially important scattering is missed. Moreover, the divergence of the cumulative gravitational forces of all the other stars (excluding the closest neighbour) at the solar mass star is also recorded, which may be used to approximate the cluster's overall potential. These will be used in the simulation of the planetary orbital evolution in Section \ref{sec-pla}.

\subsubsection{Results of the NomiClu setup}
\begin{figure*}
\includegraphics[width=1.6\columnwidth]{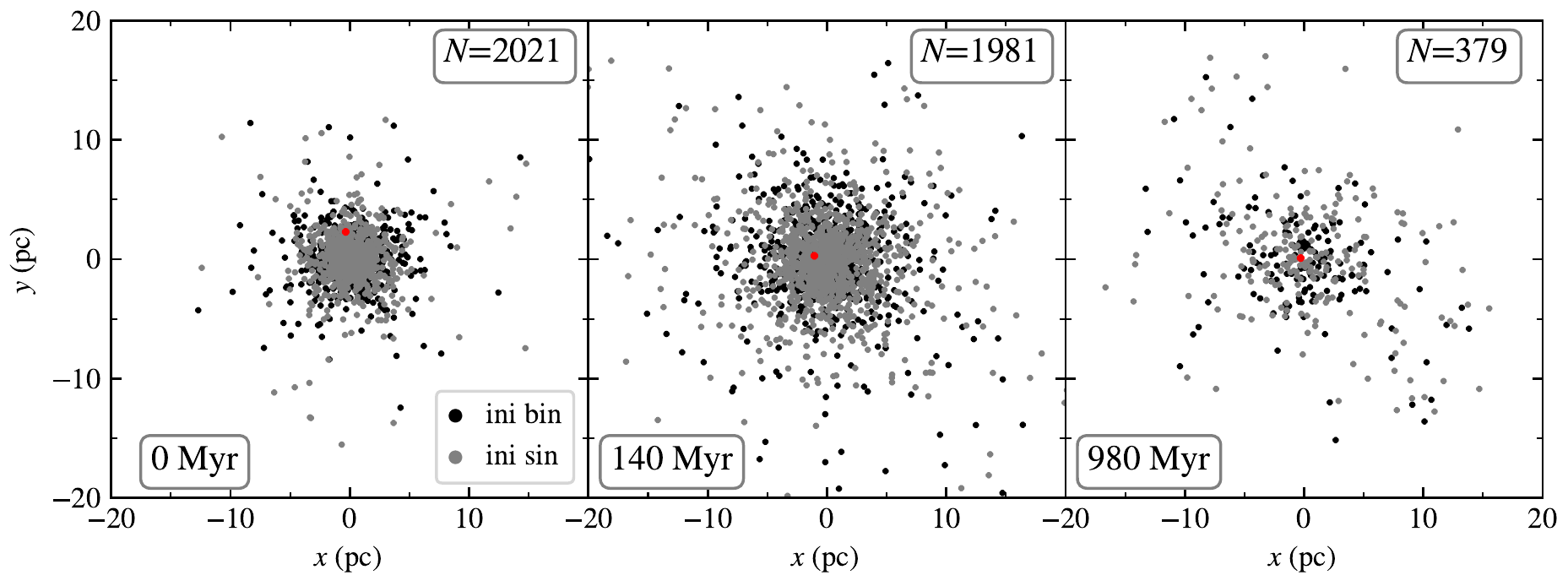}
\caption{Snapshots of a cluster in the NomiClu runs. The colour black is used to show the position of single stars and grey that of the barycentre of binaries. The red-coloured star is the one shown later in Figure \ref{fig-hjaei}, around which XZKL has been observed. The time and the number of stellar systems of the snapshots are shown at the bottom left and top right corners of each panel.}
\label{fig-clus-evo}
\end{figure*}

Figure \ref{fig-clus-evo} shows the snapshots of a run randomly selected from the NomiClu cluster realisations. The cluster is initially relatively compact but gradually expands and loses member stars. The number of stellar systems in the cluster, as shown in the top right corner, is 2021 at 0 Myr, though 2000 systems are created in {\small McLuster}. The discrepancy has to do with our binary detection method. For each star, we search for its closest neighbour and if bound, we deem the pair a valid binary. This means that we may miss the wide binaries where a third star happens to sit in between the two components.

Among the 20 initialisations of the NomiClu runs, a total of 629 solar mass stars, 246 initially single and 383 initially binary, stay in the cluster for the entire simulation of 1 Gyr. Among the former, 46 have had, at some stage during the cluster evolution, a companion for at least a complete relative orbital revolution. The fraction 19\% is in rough agreement with \cite[][though the initial condition is different]{Malmberg2007} but is considerably lower than suggested in \citet{Li2020c} where as much as 50\% was observed, a matter we discuss later in Section \ref{sec-dis-cav}. These 629 stars will be used in the planet simulation in Section \ref{sec-pla}.

\subsubsection{Statistics for all cluster setups}
\begin{figure}
\includegraphics[width=\columnwidth]{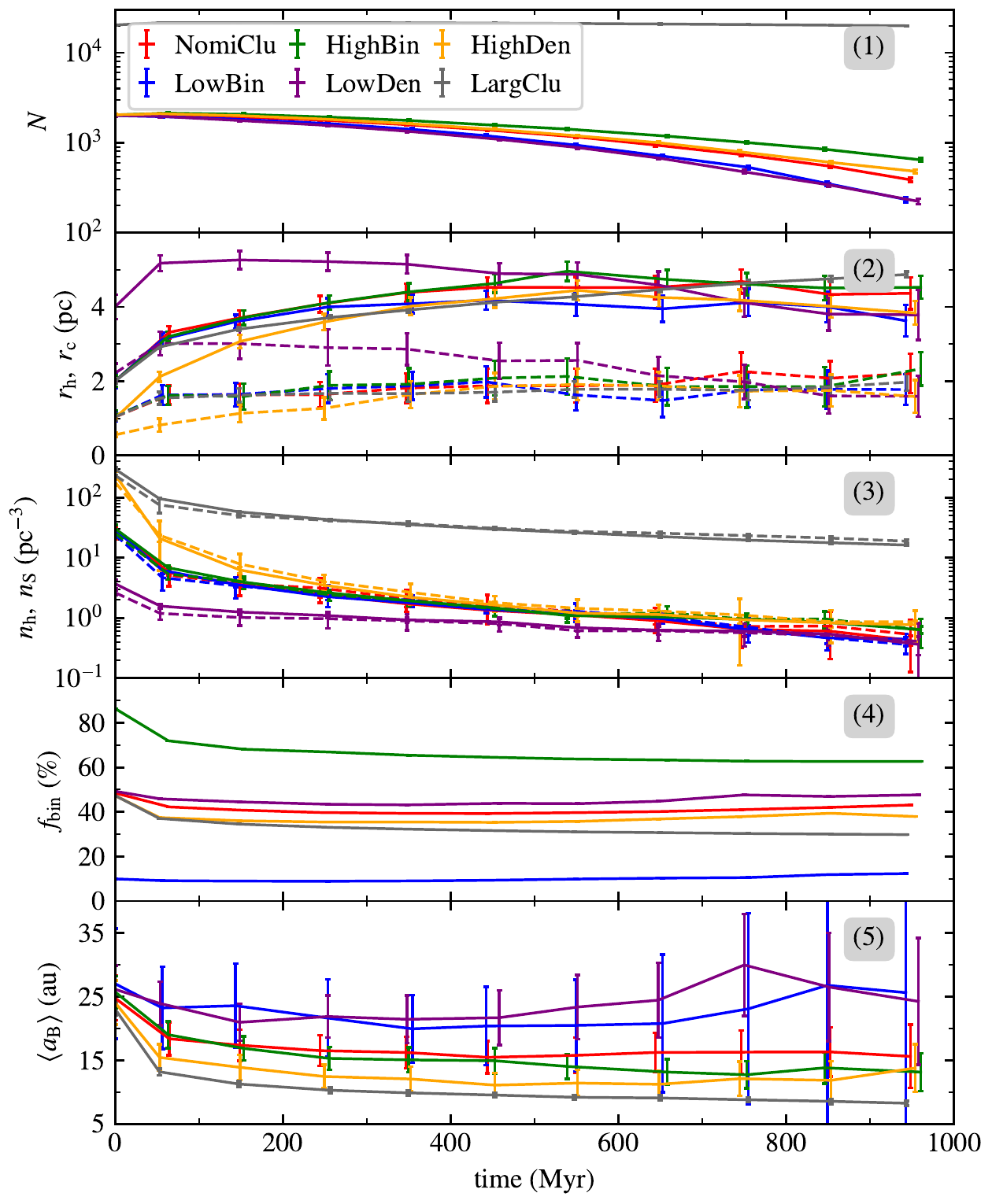}
\caption{Time evolution of the cluster properties of different models. Panel (1) shows the number of stellar systems (singles+binaries) within the cluster $N$, with different colours for different models. Panel (2) shows the half mass radius ($r_\mathrm{h}$, solid lines) and the core radius ($r_\mathrm{c}$, dashed lines). Panel (3) shows the stellar density within $r_\mathrm{h}$ ($n_\mathrm{h}$, solid lines) and the average stellar number density in the neighbourhood of all solar mass stars ($n_\mathrm{S}$, dashed lines). Panel (4) shows the clusters' binarity $f_\mathrm{bin}$ and (5) the median of the binary relative semimajor axis $\langle a_\mathrm{B}\rangle$. Error bars throughout the paper are the $1-\sigma$ scatter from bootstrapping.}
\label{fig-clu-prop}
\end{figure}

The detailed evolution of the clusters is not the focus of this work. Nonetheless, we briefly summarise how the parameters that may affect XZKL formation evolve with time in Figure \ref{fig-clu-prop}.

Panel (1) shows the total number of stellar systems in the cluster $N$, i.e., the number of single stars plus the number of pairs of binaries. For the cluster models starting with 2000 stellar systems, the number of stars drops to a few to several hundred at 1 Gyr. In the first few hundreds of Myr, one observes that $N$ somewhat increases, a result of the breakup of wide binaries. After this phase, the number of stars in all cluster models steadily decreases. At 1 Gyr, the HighBin clusters keep the largest $N$ while the models LowBin and LowDen have the least $N$. The run LargClu has yet to evolve its $N$ significantly because of its much longer evolution timescale.

Panel (2) shows the half mass radius $r_\mathrm{h}$ (solid lines) and core radius $r_\mathrm{c}$ (dashed lines), the latter being density-weighted \citep{Casertano1985}. One observes that $r_\mathrm{h}$ of all clusters asymptotically approaches $\sim4$ pc while $r_\mathrm{c}$ is about half, all cluster models within $1-2 \sigma$. For the LargClu cluster, $r_\mathrm{h}$ increases steadily reaching 5 pc at 1 Gyr while $r_\mathrm{c}$ rises initially and then fluctuates around 2 pc.

Panel (3) shows the cluster's stellar density within the half-mass radius $n_\mathrm{h}$ (solid lines) and density around solar mass stars $n_\mathrm{S}$ (dashed lines). We see that the two densities agree with each other within $1-\sigma$ across the different models. The final density is $\lesssim 1$ pc$^{-1}$ for all clusters except for the LargClu run, where the density is an order of magnitude higher.

Panel (4) shows cluster binarity $f_\mathrm{bin}$. The evolution of $f_\mathrm{bin}$ is driven by multiple factors: a binary may be destroyed by stellar scatterings and this process also affects the separation between the two binary components \citep{Heggie1975,Hills1975}; binaries are typically more massive than singles so they are more likely to sink to the centre and on the contrary, singles are more prone to ejection \citep{Hurley2007}. For most of our runs, $f_\mathrm{bin}$ remains largely unchanged (relative variation within 10-20\%, unless the initial $f_\mathrm{bin}$ is very high) and a closer examination suggests a first-decrease-then-increase trend \citep{Hurley2005}. For the HighBin runs though, $f_\mathrm{bin}$ drops significantly from 90\% to around 60\%. These trends compare well with previous simulations \citep{Kroupa1995,Hurley2007}. For the LargClu simulation, we just remind that $f_\mathrm{bin}$ may continue to evolve for many gigayears, much beyond our simulation length \citep{Hurley2007,Fregeau2009}.

Finally, the evolution of the median of the binary relative semimajor axis $\langle a_\mathrm{B}\rangle$. That of the NomiClu runs steadily declines from about 25 au to 15 au at 1 Gyr, a trend in agreement with, for example, \citet{Kroupa1995,Parker2011,Marks2012,Ballantyne2021}. Though the scatter is large, the values for the LowDen and LowBin runs have not evolved much because star scattering is weaker in these two setups. On the contrary, $\langle a_\mathrm{B}\rangle$ of the LargClu run drops the fastest to $<10$ au towards 1 Gyr as a result of strong dynamical processing in this dense cluster \citep{Parker2009}. The sets HighBin and HighDen also have $\langle a_\mathrm{B}\rangle$ declining faster than in NomiClu. How the binary's $a_\mathrm{B}$ evolves has direct consequences on XZKL, which we discuss in Section \ref{sec-pla}.

\subsection{Planets' XZKL simulation}\label{sec-pla}
\subsubsection{Simuation methods}
From the cluster simulation, we have obtained, for each solar mass star, the high-cadence time series of the mass and the state vectors of its closest neighbour, be it a companion or passerby, and the cluster potential in its vicinity. These are passed to the simulation examining the evolution of planets around the solar mass star, now performed using {\small MERCURY} \citep{Chambers1999} as we describe below.

For each solar mass star (B1, single or binary, tight or wide, Figure \ref{fig-pot-illu}), we assign 50 Jupiters as massless test particles to it, ignoring the mutual gravitational interactions between the planets; this is equivalent to simulating 50 independent single-planet systems under the perturbation of the same neighbour star at once. As in Section \ref{sec-pot-num}, the planet's initial $a=5$ au, $e=0$, and orbital normal isotropic in space; the phase angles are randomised between $0^\circ$ and $360^\circ$. We note that the possibility of XZKL does not depend on the planet's initial location sensitively \citepalias{Li2023}. Then another star (B2) is added to the system as a big body so its gravitational effect on every other body is taken into consideration. The star B2 serves as the closest neighbour (bound or not) of B1 in the parent cluster and will therefore take the mass and location and velocity vectors as recorded in Section \ref{sec-clu}. This means that these quantities may change discontinuously as the closest neighbour changes in the cluster simulation. Though, this should not have an appreciable impact on the orbital evolution of the planets, which is driven mainly by secular forcing, as we will see below.

Our test simulations show that the direct effect of the cluster potential on the planet's orbital evolution is never important because of its small orbital radius (cf., Figure \ref{fig-timescale}) so this effect is indeed turned off in the main simulations. Moreover, these tests agree with previous works that should B2 be passing by B1, it has to be very close for the planets to feel it appreciably \citep[$\ll 100$ au][]{Li2019,Li2020a}. Therefore, in our main simulations, if the current distance of B2 to B1 is larger than 2000 au, that star is omitted and the planets are moved analytically assuming two-body orbits to the point where the next star comes in.

Each system of host-planet-companion (B1, 50 planets, and B2) is run for 1 Gyr. During the simulation, we record the instance where the heliocentric of a planet is smaller than 0.022 au, roughly five times the solar radii: these planets are called XZKL-enabled. At this distance, tidal interaction between the planet and the host star is strong and may shrink and circularise the planet's orbit, leading to the formation of an HJ \citep[][\citetalias{Li2023}]{Wu2003,Fabrycky2007,Beauge2012}. But the exact outcome depends on the tidal model and in general, a fraction of those planets collide with (or get tidally disrupted by) the host star as tides cannot stop further excitation of the eccentricity and the others are turned into HJs as described above. The likelihoods of the two outcomes are by and large comparable. A stronger tidal prescription may convert into HJs some planets that are set to otherwise collide with or get disrupted by the central star. However strong the tides are, the sum of the fractions of HJ formation and collision (tidal disruption) is roughly constant \citep[][\citetalias{Li2023}, and this is the XZKL fraction in our terminology]{Petrovich2015,Anderson2016,Munoz2016}. In this work, tides are not included in the simulations, and thus we are not studying directly the tidal formation of HJs, which, as we discuss above, depends on the tidal model. Instead, we simply count the number of planets that undergo XZKL cycles, i.e. the ``candidate'' HJs. We note that not all planetary orbits having a pericentre distance smaller than 0.022 au do so owing to the secular ZKL mechanism. As we will see, a very tiny fraction is caused directly by the stellar scattering \citepalias{Li2023}, but we do not differentiate between them. In the simulation, a planet is recognised as ejected if its heliocentric distance is larger than 1000 au and is thought to have collided with the host if the distance is smaller than a solar radius. In both cases, the planet is removed from the simulation. Also, an XZKL-triggered planet remains in the simulation until it collides with the central star (or not). In our treatment, XZKL takes precedence in that an XZKL-enabled planet cannot be subject to ejection any more \citep[though ejection with the help of tidal disruption is not impossible][]{Guillochon2011}.

The main aim of this work is to explore the HJ formation channel in star clusters and compare it to the field. As we have shown in \citetalias{Li2023}, for initially-single stars, the cluster environment is likely advantageous for ZKL/XZKL as the acquisition of a companion through single-binary scattering is not efficient in the field because of the much higher scattering velocity \citep{Hut1983,Hut1983a,Fregeau2004}. However, for initially-binary stars, since a companion has been present upon the formation of the system, ZKL/XZKL may be exerted without alternation in the companion's orbit by the cluster potential or star scattering. And even potentially adverse is that in star clusters, the companion may be removed by scattering before XZKL is activated. Then indeed does the cluster help with XZKL in binary systems or not? This is where the Field set (Table \ref{tab-pop}) comes into play. From the binaries therein, we take pairs of which a component is a solar mass star. Then 50 planets are created isotropically around it as before and the host-planet-companion system is integrated for 1 Gyr, the companion orbit kept stationary and the results analysed analogously; this is to mimic XZKL in the field. We remind that this may not be a realistic representation of a field binary as galactic potential and star scattering may still be able to change the binary configuration and therefore affect the evolution of the planetary system \citep{Kaib2013}. We discuss this matter further in Section \ref{sec-dis-ZKL}.
\subsubsection{Results of the NomiClu clusters}\label{sec-xzkl-nomi}
\begin{figure}
\includegraphics[width=\columnwidth]{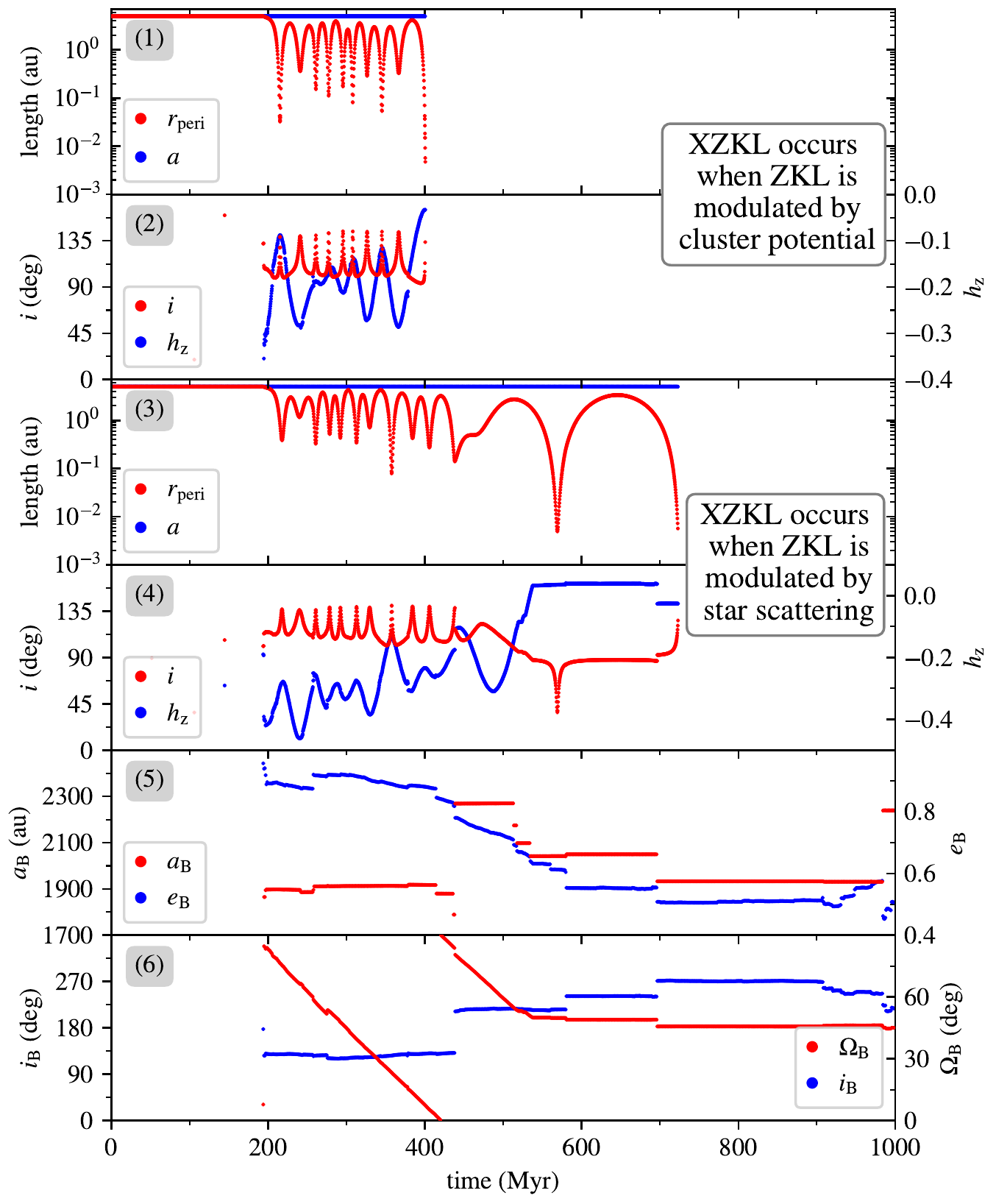}
\caption{Time evolution of the XZKL process in the numerical model. Panel (1) shows the pericentre distance of the planet's orbit $r_\mathrm{peri}$ (red) and the semimajor axis $a$ (blue); (2) shows the planet's inclination $i$ (red, left $y$-axis) and $h_\mathrm{z}$ (blue, right $y$-axis), measured against the plane of relative motion of the companion star; for this planet, XZKL occurs when the ZKL cycles are modulated mainly by the cluster potential. Panels (3) and (4) show another planet, the plot arrangement the same as the first planet; XZKL occurs when the ZKL cycles are modulated mainly by stellar scattering. Panel (5) shows the companion star's semimajor axis $a_\mathrm{B}$ (red, left $y$-axis) and eccentricity $e_\mathrm{B}$ (blue, right $y$-axis, upper panel) and (6) shows the longitude of ascending node $\Omega_\mathrm{B}$ (red, left $y$-axis) and inclination $i_\mathrm{B}$ (blue, right $y$-axis); the reference plane is that of the motion of the binary barycentre in the cluster.}
\label{fig-hjaei}
\end{figure}

We now present a detailed analysis of the XZKL phenomenon observed in the NomiClu runs. First, Figure \ref{fig-hjaei} shows the orbital evolution of two XZKL-enabled planets around the same host star that is initially single. Like the example in Figure \ref{fig-analytical}, the bottom two panels display the evolution of the companion star. Before 200 Myr, there is no data point because the star has not obtained a companion yet. At 200 Myr, the solar mass star acquires a companion on a wide ($a_\mathrm{B}\approx1900$ au) and eccentric ($e_\mathrm{B}>0.9$) orbit (Panel (5)), a feature making it (relatively) prone to orbital alternation by the cluster potential (see Figure \ref{fig-timescale}). And that is what we see in its actual evolution (Panel (6)) -- before $\sim400$ Myr, the companion is largely precessing uniformly as evidenced by the evolution of $\Omega_\mathrm{B}$; the modification in the other orbital elements is mild and there also appear instantaneous jumps as a result of strong scatterings with neighbouring stars. After 400 Myr, $e_\mathrm{B}$ drops significantly as a consequence of both secular evolution and instant changes and beyond 500 Myr, it remains around 0.5. The inclination $i_\mathrm{com}$ also instantaneously jumps from $\sim 30^\circ$ to $\sim 60^\circ$ at 420 Myr and oscillates around afterwards. Thence, the smooth precession in the binary relative orbit is seemingly quenched and there are only discontinuous jumps in the orbital elements due to stellar scatterings. Inferring from Equation \eqref{eq-tau-B}, $\tau_\mathrm{\Omega,B}$ increases when $a_\mathrm{B}$ and $e_\mathrm{B}$ decrease, both observed in the second half of the simulation in Figure \ref{fig-hjaei}. Therefore, the cluster potential-induced precession in the binary relative orbit has been weaker after 500 Myr.

In the same figure, Panels (1)  and (2 )show the temporal evolution of an XZKL-enabled planet around this star. Before the acquirement of the companion at 200 Myr, there is hardly any change in the planet's orbit. After the host star becomes a member of a binary, the planet's orbit is forced to follow periodic ZKL oscillations. Notably in Panel (2), $h_\mathrm{z}$, a constant in the standard ZKL theory \eqref{eq-pla-quad}, is evolving both continuously and abruptly, not showing any sign of conservation, as a consequence of the change in the companion's orbit caused by the cluster potential and star scattering. At 400 Myr, $h_\mathrm{z}$ approaches zero and $r_\mathrm{peri}$ dips well below 0.022 au (Panel (1)), qualifying for XZKL (and we note its ZKL is mainly modulated by the cluster potential), and the planet indeed collides with the host star and is removed.

Then Panels (3) and (4) show the evolution of another planet that is XZKL-enabled. For this planet, behaviours similar to the previous planet are observed: the planet's ZKL cycles are modulated by the cluster potential before 500 Myr. Markedly, at 520 Myr, $h_\mathrm{z}$ changes sign and the planet switches from retrograde to prograde motion (and back again at 700 Myr). After 500 Myr, the planet's ZKL cycles are only affected by the star scatterings. After an encounter at 700 Myr, the planet's $h_\mathrm{z}\sim0$, and plunges into the host. The planet becomes XZKL-enabled at 580 Myr and 730 Myr, when its ZKL is modulated mainly by stellar encounters.

From these two examples, we see that both star scattering and cluster potential can modulate ZKL cycles, by changing the orbit of the companion, in the planet's orbit and may thus give rise to XZKL. However, it is hard to tell which is more important since the two factors are often simultaneously functional as in the example above. But for a companion on a tight orbit, likely only scattering is operating as the timescale of the alteration by the cluster potential is long (Figure \ref{fig-timescale}). And we remind that the synergy between star scattering and the cluster potential is reminiscent of, for a wide binary in the field, how the relative orbit is affected by encounters with the field stars and the galactic potential, which may also impact the planetary system around the member stars \citep{Kaib2013}.

The most important question this work aims to answer is how likely is a planet subject to XZKL and turning into an HJ? And what is the HJ fraction (Equation \eqref{eq-fhj}) in star clusters? For this purpose, we need to know the giant planet occurrence rate $f_\mathrm{G}$. While that around single stars can be readily taken from the literature \citep[e.g.,][]{Cumming2008}, things become more complicated in binary systems as a companion may suppress plant formation. Generally, a relatively close companion (perhaps $\lesssim100$ au) may adversely affect $f_\mathrm{G}$ significantly while a farther one has a negligible effect \citep[e.g.,][]{Ngo2017,Hirsch2021,Moe2021}. Here we simply assume that initially-binary stars tighter than a certain threshold  $a_\mathrm{B,t}$ are planet-less and remove their contribution to XZKL (since every solar mass star is assigned 50 planets at the start) when calculating $f_\mathrm{XZKL}$. We define
\begin{equation}
\label{eq-f-XZKL}
f_\mathrm{XZKL}={1\over N_\mathrm{S}}\left({\sum_{k=1}\limits^{N_\mathrm{SP}} {N_{k,\mathrm{XZKL}} \over N_{k}}}+{\sum_{k=1}\limits^{\tilde N_\mathrm{SP}} 0}\right).
\end{equation}
In the first term in the brackets, $N_{k,\mathrm{XZKL}}$ is the number of XZKL-enabled planets around the $k$th star in our simulation and $N_{k}$ is the initial number of planets around that star, so the ratio of the two is the chance that a planet becomes XZKL-activated around that star; and the summation runs over stars that have planets (the number being $N_\mathrm{SP}$). The second term represents XZKL around stars (the number being $\tilde N_\mathrm{SP}$) that do not have planets and XZKl is impossible. And $N_\mathrm{S}=N_\mathrm{SP}+\tilde N_\mathrm{SP}$ is simply the total number of all solar mass stars. We also stress that for initially-single stars, $\tilde N_\mathrm{SP}=0$ and for initially-binary stars, $\tilde N_\mathrm{SP}$ is just the number of stars of which the companion is tighter than $a_\mathrm{B,t}$. The ejection fraction $f_\mathrm{EJEC}$ is calculated analogously.

\begin{figure}
\includegraphics[width=\columnwidth]{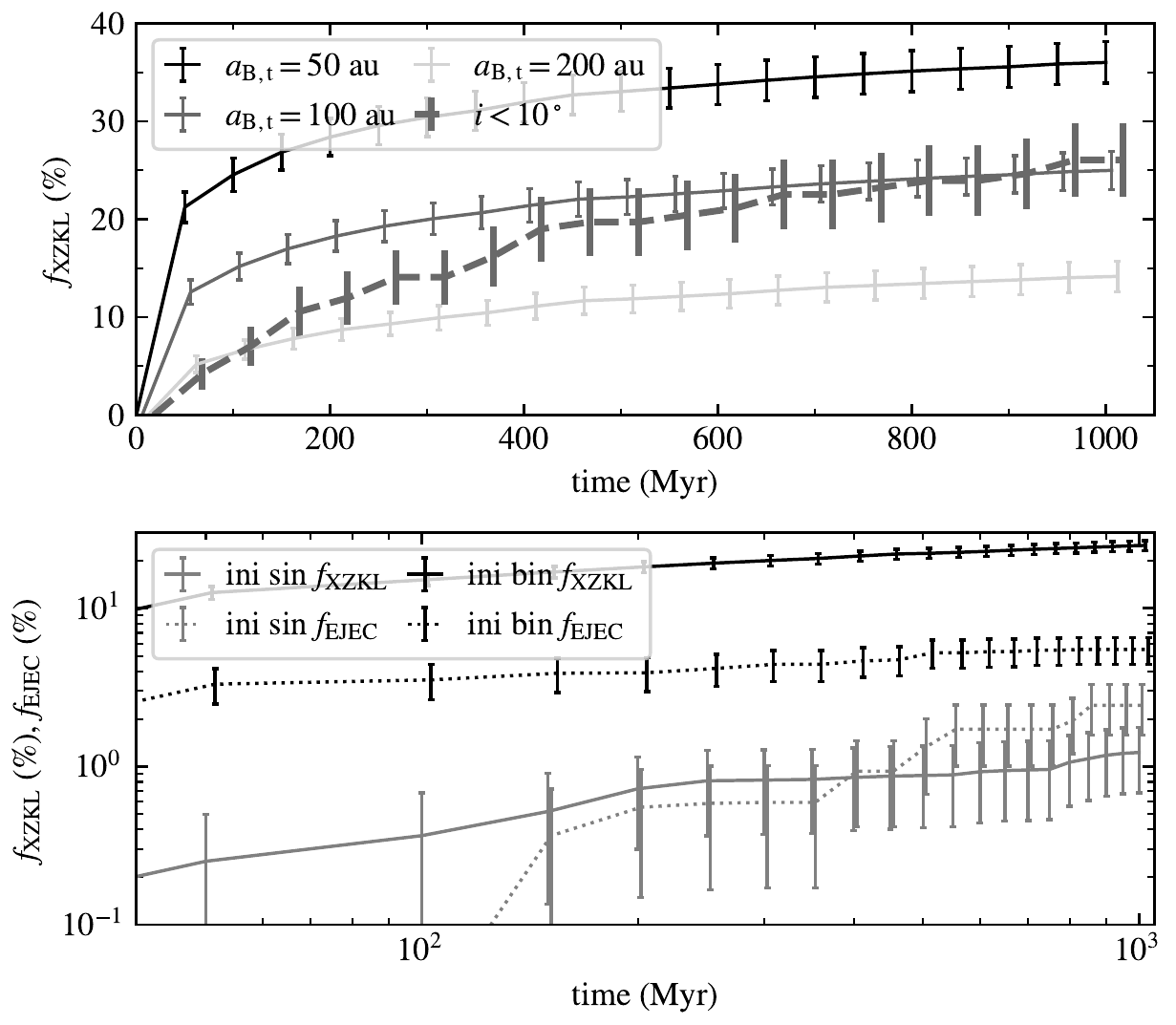}
\caption{The XZKL/ejection fraction as a function of time. Top panel: XZKL fraction $f_\mathrm{XZKL}$ \eqref{eq-f-XZKL} around solar mass stars that are initially binary. Solid lines of different greyscales show the fractions for different $a_\mathrm{B,t}$, tighter than which a member star cannot have planets. The thick grey dashed line shows the rate for initial planet-companion inclination $i<10^\circ$ and $a_\mathrm{B,t}=100$ au. The points have been shifted horizontally for better visualisation. In the remainder of the paper, we take $a_\mathrm{B,t}=100$ au. Bottom: XZKL ($f_\mathrm{XZKL}$, solid lines) and ejection ($f_\mathrm{EJEC}$, dotted lines) fraction for planets around initially binary (black lines, $a_\mathrm{B,t}=100$ au) and single (grey lines) stars.}
\label{fig-ab-cut}
\end{figure}

The thin solid lines of different greyscales in the top panel of Figure \ref{fig-ab-cut} show the evolution of $f_\mathrm{XZKL}$ for different $a_\mathrm{B,t}$ for initially-binary stars. And at 1 Gyr, $f_\mathrm{XZKL}$ reaches 36\% for $a_\mathrm{B,t}=50$ au; and those for $a_\mathrm{B,t}=100$ and 200 au approach $\sim14\%-25\%$. Notably, the differences in $f_\mathrm{XZKL}$ for the three values of $a_\mathrm{B,t}$ have been mostly established within the first 100 Myr of the simulation. This indicates that XZKL in close binary systems occurs earlier than in wide systems.

In the above, the planet's orbital normal is isotropic in space, implying that its formation is decoupled from that of the companion. However, if formed in the same disc, the two would move in the same plane. Then can XZKL still occur in this flat configuration? In order to test this, we only pick the planets of which the initial $i<10^\circ$ (measured against the binary plane of motion, and now $N_{k}$ may vary for different stars) and also with $a_\mathrm{B,t}=100$ au. The evolution of $f_\mathrm{XZKL}$ in this case is shown as the thick dashed line in the top panel of Figure \ref{fig-ab-cut} and is consistent with $f_\mathrm{XZKL}$ for $a_\mathrm{B,t}=100$ au within $1-\sigma$ during the second half of the simulation.

In the remainder of the paper, we simply apply the cut $a_\mathrm{B,t}=100$ au when calculating $f_\mathrm{XZKL}$. Before proceeding, we also note that if we disregard entirely from the statistics these tighter initial binaries and calculate $f_\mathrm{XZKL}$ as ${1\over N_\mathrm{SP}}{\sum_{k=1}\limits^{N_\mathrm{SP}} {N_{k,\mathrm{XZKL}} \over N_{k}}}$, the resulting fraction is 46\% for $a_\mathrm{B,t}=100$ au. That is to say, almost of the planets formed around initially-binary wider than 100 au stars become XZKL-activated. And we also emphasise that though initially-binary stars with $a_\mathrm{B}<a_\mathrm{B,t}$ cannot have planets themselves, they may help with XZKL around another star. For instance, they may become the companion star of a planet-hosting star and initiate XZKL in the planet's orbit or they may scatter with a host-planet-companion and by modifying the companion's orbital configuration, triggers XZKL.

The bottom panel of Figure \ref{fig-ab-cut} shows the fraction of XZKL (Equation \eqref{eq-f-XZKL}) around planets orbiting initially-single/binary stars in the black/grey solid lines. That around initially-binary stars is steadily increasing, reaching 25\% at 1 Gyr. Among those experiencing XZKL, a few per cent up to almost all \citep[][\citetalias{Li2023}]{Petrovich2015,Anderson2016,Munoz2016} of them do form HJs. Taking an exemplar value of 1/3 for the HJ/XZKL ratio, a fraction of 8\% of the planets orbiting initially-binary stars have become an HJ at 1 Gyr. The XZKL fraction for planets around initially single stars is an order of magnitude lower than for initially-binary stars. However, we also note that the error bar is large -- indeed among the 246 initially-single solar mass stars (hence a total of 246*50=12300 planets), XZKL has been only detected around 8 stars (and only 151 planets). And a single star may distort the statistics, which we discuss in Figure \ref{fig-cdf-com-hdir} below.

Perhaps quite surprisingly, the ejection fraction ($f_\mathrm{EJEC}$, black dotted line in the bottom panel of Figure \ref{fig-ab-cut}) for initially-binary stars is much smaller than the corresponding $f_\mathrm{XZKL}$. And for initially-single stars, $f_\mathrm{EJEC}$ (grey dotted line) and $f_\mathrm{XZKL}$ are consistent within $1-\sigma$. Moreover, we emphasise that most of our ejections are realised in the presence of a companion star, in agreement with \citet{Ellithorpe2022}, be it primordial or not.

Figure \ref{fig-hjaei} above suggests alterations in the companion's orbit may be significant before the planet's XZKL. Here Figure \ref{fig-cdf-com-hdir} presents, for XZKL-enabled planets around initially-single/binary solar mass stars in grey and black, the cumulative distribution function (CDF) of the angular difference between the orbital normal of the companion upon its acquirement at the host star (time 0 if initially-binary) and that at the time of the planet's XZKL activation. For the two types of host stars, more than 80\% and 40\% of the XZKL cases happen after a change of at least tens of degrees. This means that modulation by the cluster potential/stellar scattering of the ZKL in the planet's orbit is important. Furthermore, the CDF for initially-single stars shows large jumps; these are caused by efficient XZKL formation: when the companion's relative orbit is highly eccentric, the requirement on planet-companion inclination is relieved to a large degree \citep[e.g.,][]{Naoz2012,Petrovich2015}.

\begin{figure}
\includegraphics[width=\columnwidth]{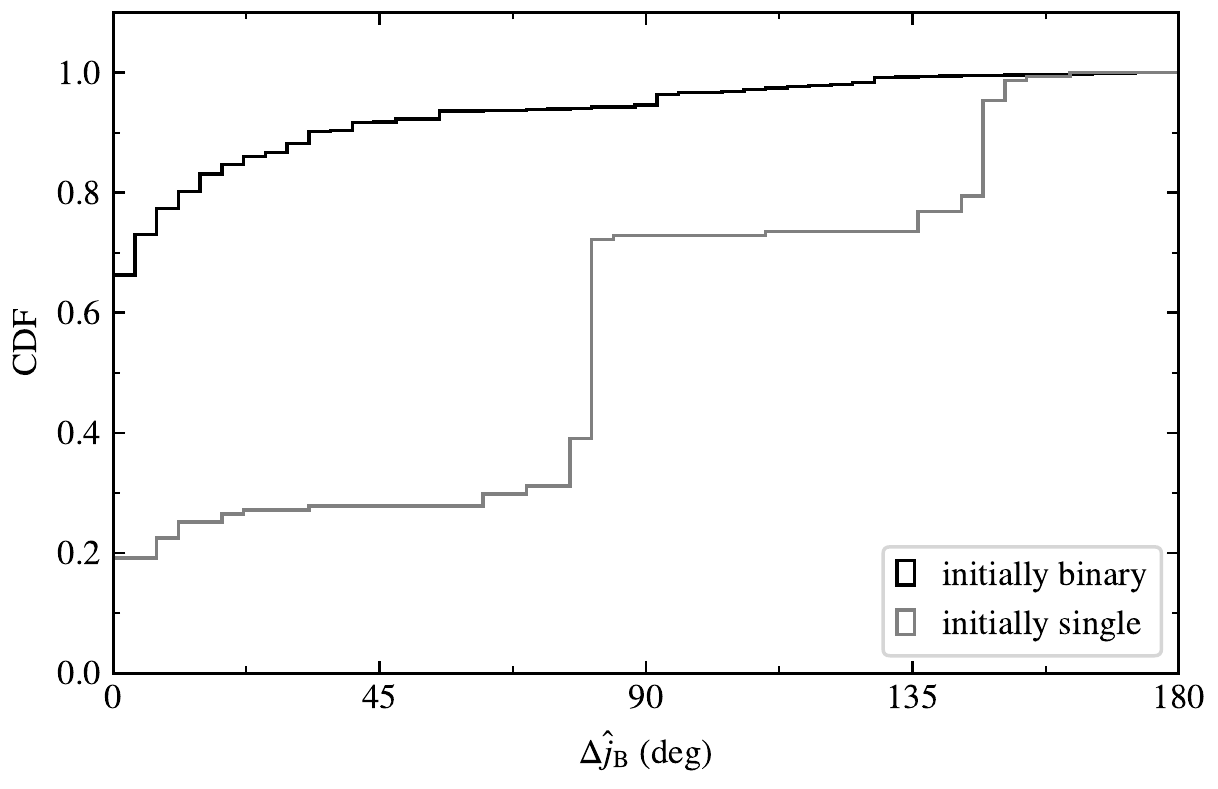}
\caption{Cumulative distribution function of the angular change in the direction of the orbital normal of the companion star. The change is measured between XZKL formation and the acquirement of the companion (start of the simulation for initially-binary stars). Each XZKL-enabled planet is counted once so a star with multiple XZKL-enabled planets is counted more than once.}
\label{fig-cdf-com-hdir}
\end{figure}

\begin{figure}
\includegraphics[width=\columnwidth]{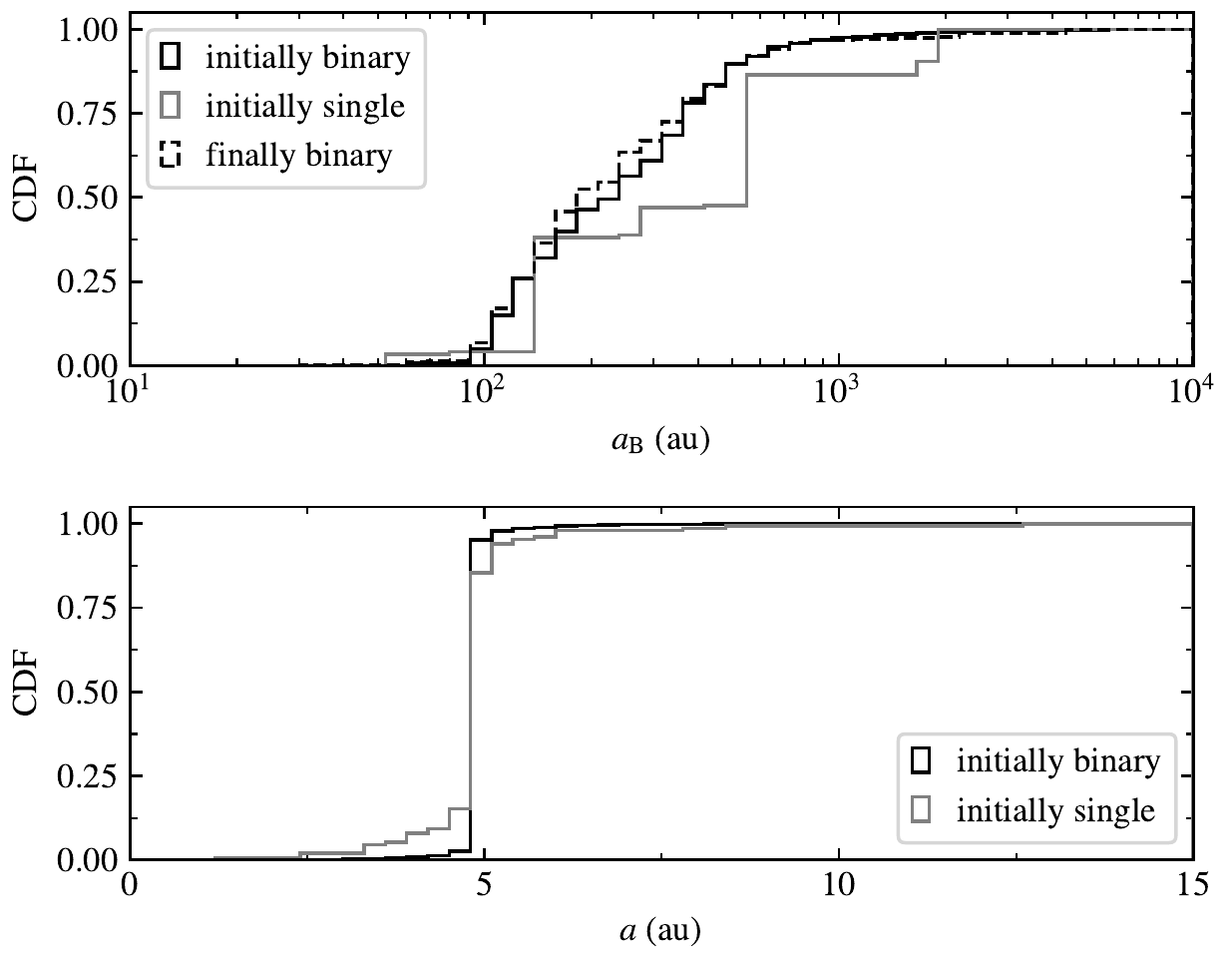}
\caption{Cumulative distribution function of the orbits of the XZKL-enabling companions and the XZKL-enabled planets. The top panel shows $a_\mathrm{B}$ of the companion star for planets from initially-binary/single stars in black/grey solid lines at the moment of XZKL; the black dashed line shows that of the finally-binary stars that host XZKL-enabled planets. The bottom panel shows $a$ of the XZKL-enabled planets from initially-binary/single stars in black/grey upon XZKL.}
\label{fig-cdf-ahj-acom}
\end{figure}

Then the top panel of Figure \ref{fig-cdf-ahj-acom} shows the CDF of the semimajor axis of the companion $a_\mathrm{B}$ the moment it activates XZKL in planets around initially-binary (black solid line) and -single stars (grey solid line). Since only initial binaries wider than 100 au are included in the analysis, there is hardly any tighter. Most of the companion orbits (initially-binary or -single) are hundreds of au wide. Less than 10\% of the companions are wider than 1000 au, for which the cluster potential may be important in changing their orbital normal. Again, for the initially-single stars, large jumps are seen. An initially-binary star may not be binary later into the cluster evolution and likewise, an initially-single star may not be single. In the same panel, the black dashed line shows the distribution of the relative of the finally-binary stars hosting XZKL-enabled planets. Compared to the CDF for initially-binary stars, there seems to be a slight shift toward tighter orbits. We discuss the observational implications in Section \ref{sec-dis-obs}.

Next, the bottom panel of \ref{fig-cdf-ahj-acom} shows the CDF of the planets' semimajor axes the moment $r_\mathrm{peri}$ dips below 0.022 au, i.e., when XZKL occurs, and planets around initially-binary and -single stars in black and grey. The huge jumps at 5 au in both curves indicate that most of these planets are only experiencing secular evolution with no significant direct scattering between the planet and neighbouring stars.

\begin{figure}
\includegraphics[width=\columnwidth]{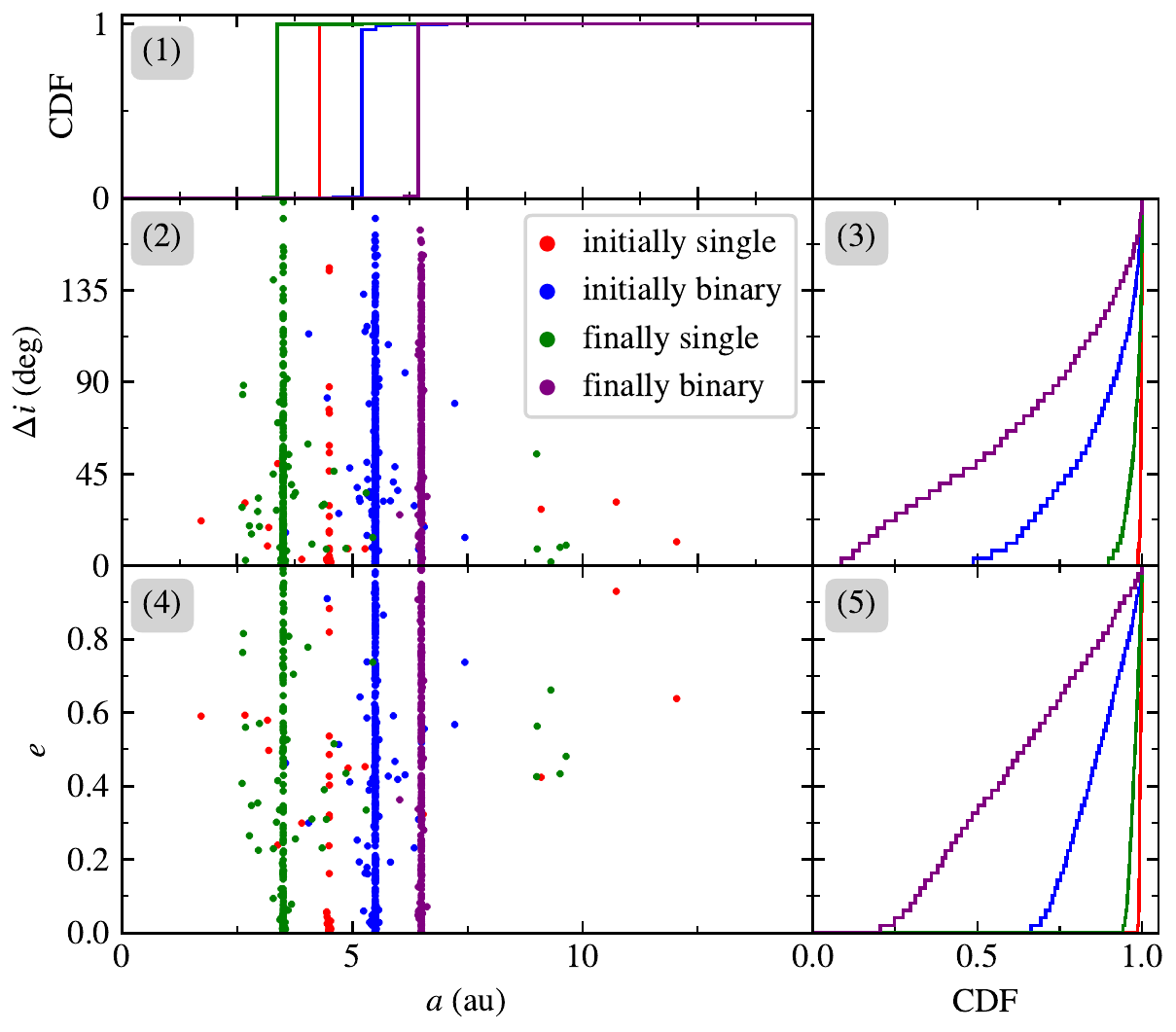}
\caption{Orbital distribution of planets surviving to 1 Gyr. Panels (2) and (4) show the distribution in the $(a,\Delta i)$ and $(a,e)$ planes where the semimajor axes of them have been shifted for clarity. The colour red/green represents planets orbiting host stars that are initially/finally single; the colours blue and purple are for planets around stars that are initially/finally binary, respectively. Panels (1), (3), and (5) show the CDF of the planets' $a$, $\Delta i$, and $e$. A planet's $\Delta i$ is the change in the direction of its orbital normal in a Gyr.}
\label{fig-dist-surv-3k}
\end{figure}

Finally, the orbital distribution of planets that survive to 1 Gyr (not subject to XZKL or ejection) is shown in Figure \ref{fig-dist-surv-3k}, divided into those from initially-single stars (red) and initially-binary stars (blue), and around finally-single stars (green) and finally-binary stars (purple). For better visibility, the semimajor axes of these groups of planets have been shifted by different amounts. Overall, the semimajor axes of all groups mostly stay close to the initial value, few wandering far. This is most obvious in Panel (1) where the CDF of $a$ is shown and again indicates that direct scattering between the planets and a star is rare.

The planets' $e$ and $i$ ($\Delta i$) are much excited. The group heated the most is the planets around finally-binary stars -- the median of $e$ and $\Delta i$ is 0.3 and $50^\circ$, respectively and more than 20\% are moving on retrograde paths (Panels (3) and (5)). We note that a planet's $\Delta i$ is the angle between its final and initial orbital normals. If the initial orbital plane coincides with the equator of the central host star, $\Delta i$ can be viewed as the stellar obliquity \citep[ignoring the evolution in the host's spin direction during ZKL,][]{Storch2014}. On the contrary, planets orbiting initially- and finally-single stars are much less excited \citep[cf.][]{Parker2012a}.
\subsection{Statistics for different cluster models}\label{sec-xzkl-diff}

\begin{figure}
\includegraphics[width=\columnwidth]{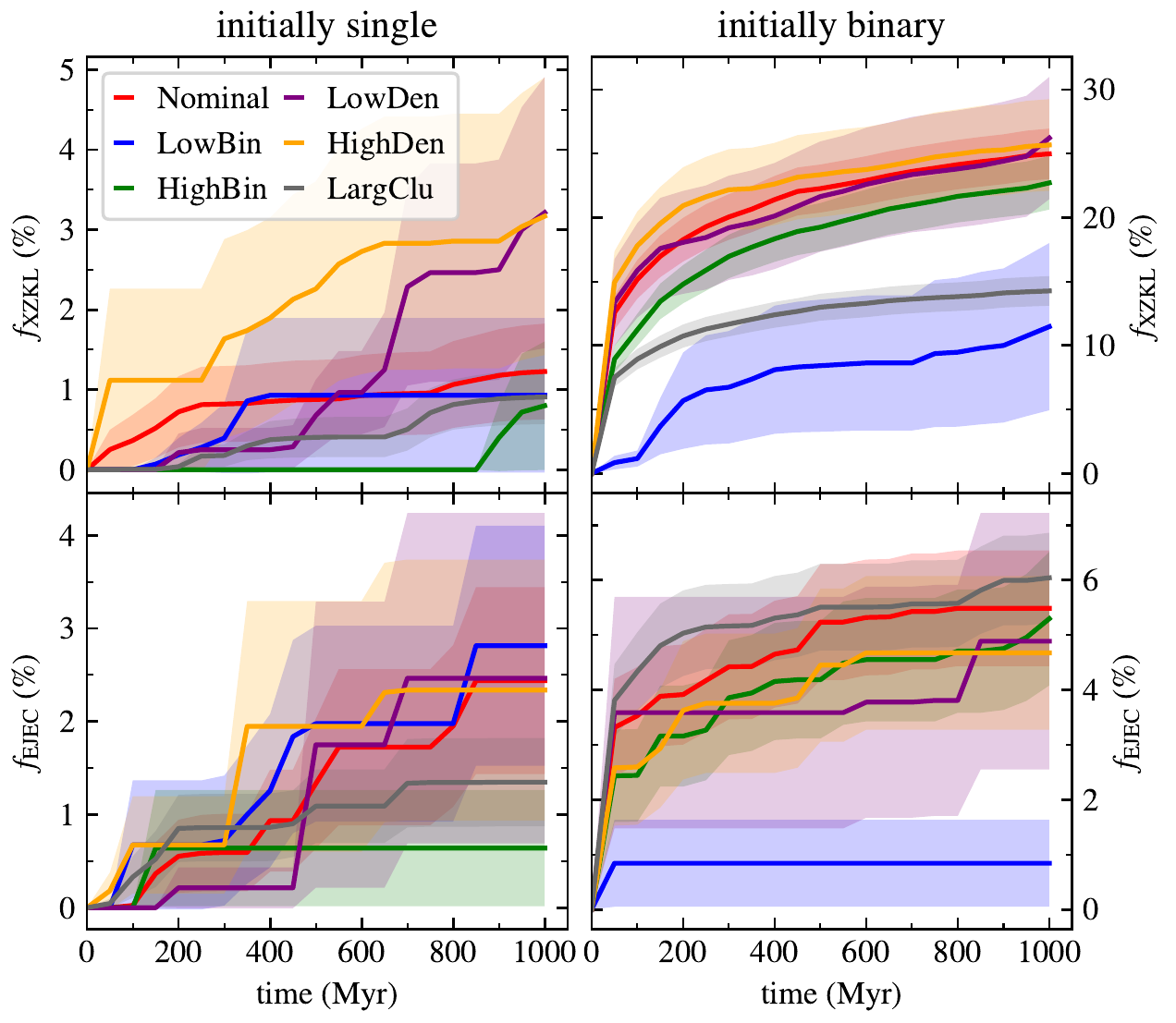}
\caption{Temporal evolution of the XZKL fraction $f_\mathrm{XZKL}$ (Equations \eqref{eq-f-XZKL}) and ejection fraction $f_\mathrm{EJEC}$ in the different cluster models. The left/right column shows those for initially-single/binary stars. The top/bottom row shows $f_\mathrm{XZKL}$/$f_\mathrm{EJEC}$. The shaded regions mark the $1-\sigma$ error. Note the $y$-scales of the panels are different.}
\label{fig-hj-cdf-all}
\end{figure}

Now we look into the XZKL formation in all of our simulations without diving into the details. Figure \ref{fig-hj-cdf-all} shows the time evolution of $f_\mathrm{XZKL}$ and $f_\mathrm{EJEC}$ for different cluster parameter sets (Equation \eqref{eq-f-XZKL}). The top left panel suggests that $f_\mathrm{XZKL}$ is a few per cent across all the cluster models. The error bars are large and essentially the results of all models agree within 1- to 2-$\sigma$. And as we have seen in Figure \ref{fig-ab-cut} already, the evolution is characterised by sudden jumps owing to efficient XZKL formation around very few stars whose companion's relative orbit is highly eccentric. Consequently, no robust link can be made between $f_\mathrm{XZKL}$ and the cluster properties from these simulations. This seems to be contradictory to \citetalias{Li2023} where a positive link between the fraction and the cluster binarity/density was constructed; we discuss this later in Section \ref{sec-dis-cav}.

For ejection around initially-single stars (bottom left panel), the fraction is a few per cent for all models and the scatter is large. This is in fair agreement with results from other cluster simulations \citep[e.g.,][though the cluster setup was different]{Fujii2019}.

The XZKL fraction $f_\mathrm{XZKL}$ around initially-binary stars (top right panel) is significantly higher, reaching 20\% or more. And again, the values from different cluster setups agree within 1- or 2-$\sigma$, except the LargClu runs and perhaps the LowBin runs also. In a large dense cluster (LargClu), binaries wider than a few hundred of au cannot survive for more than a few hundreds of Myr \citep[see also][]{Hurley2007} and the distribution of the binary semimajor therefore changes significantly \citep[e.g.,][and cf. Figure \ref{fig-timescale} and Panel (5) of Figure \ref{fig-clu-prop}]{Parker2009,Ballantyne2021}. Those much tighter may survive longer, but we simply assume that those are devoid of any planet so no XZKL is possible. Therefore, XZKL has been inefficient after $\sim 100$ Myr in this set.

The ejection fraction $f_\mathrm{EJEC}$ for planets around initially-binary stars is $\lesssim10\%$, seemingly in rough agreement with $f_\mathrm{EJEC}$ around initially-single stars and less than half of $f_\mathrm{XZKL}$ around initially-binary hosts. Notably, no obvious link can be established against the cluster property (density, binarity, size), in contrast to previous results using the concept of cross-section where a direct relationship between those quantities was constructed \citep[e.g.,][]{Li2015}.

The exact $f_\mathrm{XZKL}$ at 1 Gyr are presented in Table \ref{tab-pop}. Here, we additionally show the fraction for stars that are finally single and binary. Unlike the result for initially-single stars, here we observe $f_\mathrm{XZKL}$ for finally-single stars (from 3\% to 11\%) correlates well with the cluster binarity and density in good alignment with \citetalias{Li2023}. The LargClu is an outlier because its binary separation distribution has been substantially changed very early during the evolution (Panel (5) of Figure \ref{fig-clu-prop}). The XZKL faction for finally-binary stars is between 15\% (LargClu) and 32\% (LowDen), which seems to suggest that in the latter binary-benign case, the companion has a longer time to initiate XZKL.

\begin{table*}
\centering
\caption{Fraction of XZKL (Equation \eqref{eq-f-XZKL}) at 1 Gyr for different cluster models. The second and third columns show the fraction around stars that are initially/finally single. The fourth and fifth columns show the fractions around stars that are initially/finally in a binary. The last column displays that of all the stars.}
\label{tab-pop}
\begin{tabular}{cccccc} 
\hline
sim ID &initially single (\%)&finally single (\%) & initially binary (\%) & finally binary (\%)&overall (\%)\\
\hline
NomiClu&$1.2\pm 0.6$&$4.0\pm 0.8$&$25.0\pm 2.0$&$29.2\pm 2.4$&$15.7\pm 1.3 $\\
LowDen&$3.2\pm 1.7$&$3.9\pm 1.6$&$26.2\pm 4.4$&$32.4\pm 5.7$&$15.4\pm 2.8 $\\
HighDen&$3.2\pm 1.8$&$6.7\pm 1.9$&$25.7\pm 3.4$&$30.4\pm 4.4$&$17.1\pm 2.3 $\\
LowBin&$0.9\pm 0.9$&$1.0\pm 0.9$&$11.5\pm 6.3$&$13.0\pm 6.7$&$2.8\pm 1.4 $\\
HighBin&$0.8\pm 0.9$&$10.8\pm 2.6$&$22.7\pm 2.1$&$25.4\pm 3.0$&$20.8\pm 1.9 $\\
LargClu&$0.9\pm 0.4$&$5.2\pm 0.6$&$14.3\pm 1.0$&$14.6\pm 1.6$&$7.9\pm 0.6 $\\
Field&0&0&$5.6\pm 0.5$&$5.6\pm 0.5$&$3.9\pm 0.4 $\\
\hline
\end{tabular}
\end{table*}

The last column of Table \ref{tab-pop} shows the overall XZKL fraction in each type of cluster. It is the lowest (3\%) for the LowBin cluster model and highest (21\%) for the HighBin model. The other cluster models with 2000 stellar systems are all around 15\%. For these, the most important factor affecting the XZKL is the cluster binarity. And it is not that the XZKL efficiency is hugely different in different clusters but simply that binaries are much more capable of producing XZKL than single stars in general so when the former make up a larger fraction of the cluster stars, the overall XZKL fraction is higher. The LargClu cluster has an $f_\mathrm{XZKL}$ of only 8\% since the wide binaries favourable for XZKL have been dissociated quickly.

Now we are ready to comment on the overall effect of the cluster environment on XZKL/HJ formation. For a planet around an initially-single star, the host cannot obtain a companion star via scattering with a field binary because of the much higher relative velocity \citep{Hut1983,Hut1983a,Fregeau2004}. Therefore, XZKL is never possible (Field runs, the last row in Table \ref{tab-pop}) and the cluster environment apparently has a positive impact on XZKL/HJ formation for those single stars. For a planet around an initially-binary star, the companion may naturally excite XZKL in the planet's orbit without modification to its orbit and Table \ref{tab-pop} tells that $f_\mathrm{XZKL}$ is 6\% which is much smaller than the corresponding values in clusters. Overall, except for very low binarity clusters (the reason again being merely a number effect and we note $f_\mathrm{bin}=50\%$ for the field runs), the XZKL fraction in clusters is significantly higher than that of the field.

\section{Discussion}\label{sec-dis}

\subsection{Caveats}\label{sec-dis-cav}

In this work, only the Newtonian gravity is taken into consideration. We have not modelled the actual tidal formation and ensuing evolution of HJs \citep{Hut1981,Rasio1996a}. For a planet in XZKL, the distortion in the planet may cause the circularisation of its orbit, during which the orbital angular momentum is largely conserved. After this phase, the stellar tide is at work and there can be substantial exchanges between the orbital angular momentum and that associated with the star's rotation. As a result, the planet migrates either inwards to the star or outward, a process further modifies the HJ population \citep[][]{Jackson2009,Hamer2019,Yee2019}. This means that a fraction of our early-produced HJs will be destroyed later. However, as Figure \ref{fig-hj-cdf-all} shows, HJ creation in a star cluster, though at a slower rate, continues late into the cluster evolution and these late-produced may be observable.

For initially-binary stars, only those with a companion separation wider than 100 au are assumed to be planet hosts in our calculation. It has been suggested that perhaps there exists an enhancement for wider pairs \citep{Duchene2010,Hirsch2021}. And we remind even if the planet and the primordial companion star form in the same plane (thus unfavourable for XZKL), the XZKL fraction presented in Table \ref{tab-pop} will not be affected much (cf. Figure \ref{fig-ab-cut}).

We note that in the cluster simulations, one can specify a variable GMIN in {\small NBODY6++}. If the gravity of all the other stars compared to the binary internal forcing is smaller than GMIN, the binary would be treated as if it is in isolation (the barycentre moves in the cluster but no evolution in the relative orbit) in order to save the computer time. This occurs when the binary is not close to any other stars (no stellar scattering) and the cumulative gravity of the distant stars (cluster potential) is weak. Thus, this might affect our ability to model consistently the effect of the cluster potential on XZKL. The GMIN value we have used in our simulations is $10^{-7}$. Our test runs show that in different clusters, this choice would mean that for binaries tighter than $\sim100-300$ au, the influence of the cluster potential may not be accounted for. But this only has a minor effect on XZKL since (1) according to Figure \ref{fig-timescale}, the cluster potential is only important for very wide binaries and (2) binaries tighter than 100 au do not contribute towards XZKL.

Moreover, we have used only the closest neighbour in the parent cluster of the planet-hosting stars during the planets' XZKL simulations. While probably not affecting the secular dynamics of the ZKL cycles of the planet, this approach makes it impossible to account for consistently the perturbation to the planet when the planetary system is scattering with a binary star. Compared to the scattering with a single star (which is modelled faithfully in our planet simulation), that with a binary is several times more destructive \citep{Laughlin1998,Li2015,Li2020c,Wang2020}. As shown by \citet{Li2020c}, in an open cluster of $f_\mathrm{bin}=50\%$, the chance for a planetary system (planet at 5 au) to acquire a companion star (planet intact) is seven times as likely as the planet's ejection during star scatterings. The former possibility has been measured by our simulations and is 20\%. Hence perhaps 3\% of the planets are removed during star scatterings, a process not fully captured in our simulation. Hence the effect scattering with binary stars on XZKL fraction is probably very minor. 


In our cluster simulations, all setups have adopted a smooth initial distribution. An open cluster may be born with substructures. Though the substructures themselves dissolve quickly \citep{Parker2012}, the temporary high density regions may dissociate wide binary stars very efficiently in moderate clusters \citep{Parker2011}, perhaps like observed in our LargClu run \citep{Parker2009}. This would likely affect XZKL/HJ formation adversely.

Our preceding work \citetalias{Li2023} concentrated on HJ formation around initially-single stars under the effect of star scattering only and found a positive dependence of the HJ fraction on the cluster binarity and density. Therein, a static cluster model with fixed stellar density, binarity, and binary orbital distribution has been adopted. Here in our more realistic simulations, these quantities naturally change as the cluster is consistently evolved using an $N$-body code. Notably, both stellar density and the binary orbits change more dramatically in initially denser and more binary-rich clusters (Figure \ref{fig-clu-prop}), which adversely affects XZKL. As a result, the link between HJ fraction and cluster property constructed in that work is in some sense invalidated: for instance, the densest setup LargClu now has the smallest XZKL/HJ fraction.

As \citetalias{Li2023}, here only a single-planet system is modelled. While the orbital semimajor axis and eccentricity of the planet will not statistically affect the XZKL/HJ fraction \citepalias{Li2023}, additional planets will add to the complexity of the system under the perturbation of a companion star. On the one hand, the planetary system may be destroyed, leading to usually the ejection of one or more planets and leaving the survivors on much-excited orbits \citep{Malmberg2007a}. On the other, the planetary system may hold together by the interplanetary gravity and precesses as a rigid body \citep{Innanen1997}. In both scenarios, the XZKL/HJ fraction likely deviates from the prediction for single-planet systems in this work.
\subsection{Modification to ZKL effects on a planetary system}\label{sec-dis-ZKL}
We show in this work that in the cluster-binary-planet system, the ZKL forced in the planet's orbit by the companion can be modulated by the cluster either by its gravitational potential or by direct scattering by altering the companion's orbit. This ``modulated'' ZKL phenomenon has been studied elsewhere.

Perhaps the one carrying the most analogy is \citet{Kaib2013}. Those authors studied how in our Galaxy, a binary's relative orbit may be subject to modifications by star scatterings and the galactic potential and how this affects the planetary system around a member star. They found that the companion pericentre may dip toward the planetary system, triggering instability and the resultant distribution in the surviving planet's eccentricity matching the observations of exoplanets in very wide binaries.

A few other works have also examined the evolution of the orbits of binary stars in clusters and discussed how this affects planetary systems around a component star. \citet{Parker2009a,Ballantyne2021} studied how the primordial binaries' relative orbits are processed and showed that within 10 Myr $\sim10\%$ of them would have been turned by more than $40^\circ$, therefore potentially able to trigger ZKL in the planets' orbit (but the planet evolution has not been simulated). \citet{Ellithorpe2022} explicitly explored how a planetary system with a primordial coplanar companion may be susceptible in a cluster. For the solar system, they found roughly 10\% of the systems are destabilised, likely caused by direct scattering with the companion whose orbit is modified by stellar encounters. In all these works, the length of the integration has been no more than 10 Myr. Hence it is unlikely that the alternation of the companion orbit by the cluster potential can be manifested (Figure \ref{fig-timescale}); also, planets around initially single stars have been omitted and HJ formation not the focus.

Analogous to the cluster/galactic potential, an additional body may also modulate the ZKL cycles exerted on the planet's orbit. \citet{Hamers2017b} shows that a distant star orbiting the host-planet-companion can increase the HJ formation rate by a few tens of per cent and dramatically boost the rate of tidal disruption of the planet close to its host.

Furthermore, from within the star-planet-companion system, short-range forces, like general relativity, tides, and non-spherical-shape that operate after XZKL has been activated, may cease further excitation and lead to the formation of HJ \citep{Wu2003,Liu2015}. The statistics have been looked at by \citet{Fabrycky2007,Wu2007,Naoz2012,Petrovich2015,Anderson2016,Munoz2016,Vick2019} and it was found that in general the fraction of XZKL most crucially depends on the system's orbital configuration.

Finally, if the planet's host star is a relatively massive one, it may evolve off the main sequence, lose a fraction of its mass and become a white dwarf. During the mass shedding, the planet's and companion's orbits may adiabatically expand at different rates and stronger ZKL may be excited in the planet's (or a small body's) orbit \citep{Stephan2017}.
\subsection{Observational implications}\label{sec-dis-obs}

First, we estimate the HJ fraction in the open clusters we simulate. As discussed in Section \ref{sec-xzkl-nomi} already, how an XZKL-activated planet turns into an HJ is not modelled explicitly. We simply assume that 1/3 of XZKL-enabled planets become HJs \citep[$f_\mathrm{HJ/XZKL}=1/3$,][\citetalias{Li2023}]{Petrovich2015,Anderson2016,Munoz2016}. Then our predicted HJ fraction $f_\mathrm{HJ}$ follows
\begin{equation}
\begin{aligned}
\label{eq-fhj}
f_\mathrm{HJ}=& f_\mathrm{HJ/XZKL}\times f_\mathrm{XZKL}\times f_\mathrm{G}\\
=&{1\over3}\times15\%\times10\%\sim0.5\%
\end{aligned}
\end{equation}
where $f_\mathrm{G}$ is the occurrence rate of a giant planet at several au around a solar mass star and $f_\mathrm{XZKL}$ assumes a value of 15\% as from Table \ref{tab-pop} (so the cluster binarity is not too small and the number density not too high). Observationally, $f_\mathrm{G}$ is about 10\% for field stars \citep[][and more recently, \citealt{Fernandes2019,Wittenmyer2020,Fulton2021}]{Cumming2008,Mayor2011}. Unless the cluster is extremely dense, the protoplanetary disc is not affected much \citep[e.g.,][]{Adams2006,Winter2018} so we assume that $f_\mathrm{G}=10\%$ for our clusters as well. Then our model predicts an $f_\mathrm{HJ}$ of 0.5\% at 1 Gyr.

M67 possibly formed with an initial configuration like that for our LargClu run, but with an initial half-mass radius twice as large \citep{Hurley2005}. Therefore, the initial stellar density in M67 is not as high as our LargClu cluster but is similar to the NomiClu runs. Moreover, M67 is about solar age, so more HJs would have formed (after 1 Gyr) compared to Equation \eqref{eq-fhj}. Then the HJ fraction of M67, as deduced from our mechanism, is perhaps 0.01, still short of the observed value of 0.05 \citep{Brucalassi2016} by a factor of a few.

Our results show that an open cluster's HJ fraction is affected by its age, binarity in a positive way, and density in a negative manner. We advocate for HJ surveys targeting old, high-binarity, not-too-dense clusters where we find HJ fraction comparably high, and for the prioritisation of wide binaries where most of the Hjs are formed. Also, we remind that the extent of heating in the planets' orbits is distinct in single and binary stars. 


Free-floating planets (FFPs) have been observed in both young Myr and relatively old 100 Myr open clusters \citep{ZapateroOsorio2000,ZapateroOsorio2014}. In our simulations, $f_\mathrm{XZKL}\sim f_\mathrm{EJEC}$ (Figure \ref{fig-hj-cdf-all}) so HJs and FFPs are created at a similar rate. But how the two populations are retained \citep[e.g.,][]{Rasio1996a,Hurley2002a,Jackson2009,Liu2013} needs further detailed discussion. But we point out that observational constraints on one population may have implications on the other.

Finally, we remind that in this work, we have examined only the formation of HJs around stars that remain in the cluster. During the cluster evolution, the member stars continue to escape, turning into field stars. Among those are HJ-hosting stars (cluster-HJ) and they become part of the field-HJ population. To account for the contribution of cluster-HJs to field-HJs needs more careful modelling of the star-planet interaction and is left for future study.

\section{Conclusions}\label{sec-con}
We have looked into HJ formation via XZKL tidal migration in star clusters. First, in the cluster-binary-planet system, we use secular theory to study the (X)ZKL cycles in a planet's orbit where the binary relative orbit is modulated by the cluster potential. Then we perform full-fledged Gyr $N$-body simulations to examine XZKL in a more realistic cluster environment. This is divided into two steps: the first deals with the cluster evolution and information on the companion/neighbour stars of planet-hosting stars is fed to the second stage where the evolution of the planetary system, under the perturbation of the companion/neighbour, is explored. The star cluster's size, density and binarity are varied while all planets are initially at 5 au from the host. Our main findings are as follows.
\begin{itemize}
\item The relative orbit (eccentricity, inclination, and phase angles) of a stellar binary evolves secularly owing to the cluster's gravitational potential. For binaries on very wide and eccentric orbits, the timescale of this effect may be shorter than other dynamical processes.
\item If one of the binary components hosts a planet, the alteration in the binary relative orbit by the cluster potential may change the host-planet-companion configuration from an XZKL-implausible regime to a favourable one.
\item Our cluster simulation of different initial cluster properties shows that while the star number and density decline with time, the binarity remains largely unchanged.
\item An initially-single planet-hosting star may, during the cluster evolution, acquire a companion via scattering with a binary, which may excite XZKL cycles in the planet's orbit and the XZKL fraction is $f_\mathrm{XZKL}\sim0.8\%-3\%$ at 1 Gyr (see Equation \eqref{eq-f-XZKL}).
\item Around stars that are initially binary, $f_\mathrm{XZKL}\sim12\%-26\%$ and in the vast majority of these, the XZKL is driven by the primordial companion and for the remainder, a new companion star formed via stellar scattering has been at work.
\item For the two types of stars, alteration in the companion orbit is important: a substantial fraction of XZKL occurs after significant changes (tens of degrees) in the orbital normal of the companion due to star scattering and cluster potential.
\item Around initially-single stars, the ejection fraction $f_\mathrm{EJEC}$ is a few per cent or less, comparable to XZKL. Around initially-binary stars, $f_\mathrm{EJEC}$ is also a few per cent, in this case substantially smaller than $f_\mathrm{XZKL}$.
\item Around the stars that are finally binary, $f_\mathrm{XZKL}\sim13\%-32\%$ and around finally-single stars, the fraction is roughly $f_\mathrm{XZKL}\sim1\%-11\%$.
\item Combined, the overall $f_\mathrm{XZKL}$ is 3\%-21\% at 1 Gyr which depends positively on the cluster binarity. And in a larger denser cluster, $f_\mathrm{XZKL}$ is indeed lower where the XZKL-favourable binary configurations are more vulnerable to star scattering.
\item Though the cluster environment significantly boosts the HJ formation rate (a factor of four comparing the NomiClu and the Field simulations), the efficiency is not enough to match the observations of M67.
\end{itemize}
\section*{Acknowledgements}
The authors acknowledge financial support from the National Natural Science Foundation of China (grants 12103007 and 12073019) and the Swedish Research Council (grant 2017-04945) and the Swedish National Space Agency (grant 120/19C) and the Fundamental Research Funds for the Central Universities (grant 2021NTST08).

This work has made use of {\small NumPy} \citep{Harris2020}, {\small SciPy} \citep{Virtanen2020}, and {\small Matplotlib} \citep{Hunter2007}.

\section*{Data Availability}
The data underlying this paper will be shared on reasonable request to the corresponding author.



\bibliographystyle{mnras}



\appendix

\section{Derivation of the secular equations of motion}

The potential of the cluster at B1/B2 can be obtained by substituting the quantity $r$ with $r_1$/$r_2$ in Equation \eqref{eq-cluster-pot} (the last two terms on the right-hand side of Equation \eqref{eq-tot-eng}). Then $r_1$ and $r_2$ can be expressed using $r_\mathrm{B}$, the distance from the cluster centre to the centre of mass of the binary; see Figure \ref{fig-pot-illu}. For example, $r_2$ fulfils the relation $r_2^2=r_\mathrm{B}^2+R_2^2-2r_\mathrm{B}R_2\cos\gamma$ where $R_2$ measures the distance from the binary barycentre to B2 and $\gamma$ is the angle between $\boldsymbol r_\mathrm{B}$ and $\boldsymbol R_\mathrm{B}$, the vector point from B1 to B2. Then the potential at B2 can, after some algebraic manipulations, be written as
\begin{equation}
\Phi_2=-{GM \over \sqrt{r_\mathrm{B}^2+b^2}}{m_2\over\sqrt{1+2{r_\mathrm{B}\over \sqrt{r_\mathrm{B}^2+b^2}}{R_2\over\sqrt{r_\mathrm{B}^2+b^2}}\cos\gamma+({R_2\over \sqrt{r_\mathrm{B}^2+b^2}})^2}}.
\end{equation}
The second factor on the right-hand side can be readily expanded as Legendre polynomials in $R_2/ \sqrt{r_\mathrm{B}^2+b^2}$. A similar operation can be done for the potential at B1. Then the sum of the cluster potential energies of B1 and B2 can be converted to, keeping term up to the third order of the expansion
\begin{equation}
\begin{aligned}
\Phi_{12}=&-{GM(m_1+m_2) \over \sqrt{r_\mathrm{B}^2+b^2}}\\
&-{GMm_1m_2 r_\mathrm{B}^2R_\mathrm{B}^2\over 2(m_1+m_2)(r^2+b^2)^{5/2}}(3\cos^2\gamma-1-b^2/r_\mathrm{B}^2).
\end{aligned}
\end{equation}
The first term on the right-hand side dictates the motion of the binary barycentre in the cluster and the second determines how the binary relative orbit evolves.

The binary orbital period is much shorter than that of the barycentre's motion in the cluster so the former can be averaged out \citep[e.g.,][]{Valtonen2006,Tremaine2014} and we arrive at Equation \eqref{eq-bin-pot}.

The equation of motion for ${\boldsymbol e}_\mathrm{B}$ is \citep{Tremaine2014}
\begin{equation}
\label{eq-eq-eb}
\begin{aligned}
{\mathrm{d} {\boldsymbol e}_\mathrm{B}\over \mathrm{d}t}=&-{1\over L_\mathrm{B}}({\boldsymbol e}_\mathrm{B}\times \nabla_{{\boldsymbol j}_\mathrm{B}}\phi_\mathrm{B}+{\boldsymbol j}_\mathrm{B}\times \nabla_{{\boldsymbol e}_\mathrm{B}}\phi_\mathrm{B})\\
=&{C_\mathrm{B}\over L_\mathrm{B}}[{-6({\boldsymbol j}_\mathrm{B} \cdot \hat r)\boldsymbol e}_\mathrm{B}\times\hat r-12{\boldsymbol j}_\mathrm{B}\times{\boldsymbol e}_\mathrm{B}-6{b^2\over r^2}{\boldsymbol j}_\mathrm{B}\times{\boldsymbol e}_\mathrm{B}\\
&+30({\boldsymbol e}_\mathrm{B} \cdot \hat r){\boldsymbol j}_\mathrm{B}\times\hat r]
\end{aligned}
\end{equation}
where
\begin{equation}
\label{eq-lb}
L_\mathrm{B}={m_1m_2 \over m_1+m_2}\sqrt{G (m_1+m_2)a_\mathrm{B}}
\end{equation}
is the momentum conjugate to the binary relative orbital motion \citep{Tremaine2014,Liu2015,Petrovich2015}. Likewise, the equation of motion for ${\boldsymbol j}_\mathrm{B}$ is \citep{Tremaine2014}
\begin{equation}
\label{eq-eq-jb}
\begin{aligned}
{\mathrm{d} {\boldsymbol j}_\mathrm{B}\over \mathrm{d}t}&=-{1\over L_\mathrm{B}}({\boldsymbol e}_\mathrm{B}\times \nabla_{{\boldsymbol e}_\mathrm{B}}\phi_\mathrm{B}+{\boldsymbol j}_\mathrm{B}\times \nabla_{{\boldsymbol j}_\mathrm{B}}\phi_\mathrm{B})\\
&={C_\mathrm{B}\over L_\mathrm{B}}[30({\boldsymbol e}_\mathrm{B} \cdot \hat r){\boldsymbol e}_\mathrm{B}\times\hat r -6({\boldsymbol j}_\mathrm{B} \cdot \hat r){\boldsymbol j}_\mathrm{B}\times\hat r ]
\end{aligned}
\end{equation}

In the derivation of Equation \eqref{eq-bin-pot}, the binary relative motion has been eliminated. We can further remove the motion of the binary barycentre in the cluster if it follows a circular path and the resulting doubly-averaged potential energy is
\begin{equation}
\label{eq-bin-pot-ave}
\phi_\mathrm{B}={C_\mathrm{B}\over2}[1-6e_\mathrm{B}^2+{b^2\over r_\mathrm{B}^2}(4+6e_\mathrm{B}^2)+15({\boldsymbol e}_\mathrm{B} \cdot \hat n_\mathrm{B})^2-3({\boldsymbol j}_\mathrm{B} \cdot \hat n_\mathrm{B})^2].
\end{equation}
where $\hat n_\mathrm{B}$ is the unit vector parallel to the normal of the plane of motion of the binary barycentre. Equation \eqref{eq-bin-pot-ave} can also be expressed in the normal orbital elements, leading to Equation \eqref{eq-bin-pot-ave-ele}.

The equations of motion of the planet's orbital elements can be similarly derived from the potential \eqref{eq-pla-quad}
\begin{equation}
\label{eq-eq-e}
\begin{aligned}
{\mathrm{d} {\boldsymbol e}\over \mathrm{d}t}=&-{1\over L}({\boldsymbol e}\times \nabla_{{\boldsymbol j}}\phi_\mathrm{P}+{\boldsymbol j}\times \nabla_{{\boldsymbol e}}\phi_\mathrm{P})\\
=&{C_\mathrm{P}\over L}[{6({\boldsymbol j} \cdot \hat n_\mathrm{B})\boldsymbol e}\times\hat n_\mathrm{B}+12{\boldsymbol j}\times{\boldsymbol e}-30({\boldsymbol e} \cdot \hat n_\mathrm{B}){\boldsymbol j}\times\hat n_\mathrm{B}]
\end{aligned}
\end{equation}
where
\begin{equation}
C_\mathrm{P}={Gm_2a^2\over 8a_\mathrm{B}^3(1-e_\mathrm{B}^2)}
\end{equation}
and
\begin{equation}
L=\sqrt{Gm_1a}.
\end{equation}
and
\begin{equation}
\label{eq-eq-j}
\begin{aligned}
{\mathrm{d} {\boldsymbol j}\over \mathrm{d}t}&=-{1\over L}({\boldsymbol e}\times \nabla_{{\boldsymbol e}}\phi_\mathrm{P}+{\boldsymbol j}\times \nabla_{{\boldsymbol j}}\phi_\mathrm{P})\\
&={C_\mathrm{P}\over L}[-30({\boldsymbol e} \cdot \hat n_\mathrm{B}){\boldsymbol e}\times\hat n_\mathrm{B} +6({\boldsymbol j} \cdot \hat n_\mathrm{B}){\boldsymbol j}\times\hat n_\mathrm{B} ].
\end{aligned}
\end{equation}

\bsp	
\label{lastpage}
\end{document}